\documentclass{article}

\usepackage{graphicx}  
\usepackage{amsmath}
\usepackage{amssymb}   
\usepackage{amsfonts}
\usepackage{subcaption}
\usepackage[utf8]{inputenc}
\usepackage{algorithm} 
\usepackage{algpseudocode}
\usepackage{enumitem}
\usepackage{tikz} 
\usepackage{hyperref}
\usepackage{accents} 

\usepackage{authblk}

\newif\ifcomments
\commentsfalse
\commentstrue 

\ifcomments
\newcommand{\mmm}[1]{{\color{red}\textbf{MM: #1}}}
\newcommand{\jmh}[1]{{\color{blue}\textbf{JMH: #1}}}
\newcommand{\pjj}[1]{{\color{magenta}\textbf{PJ: #1}}}
\newcommand{\lc}[1]{{\color{cyan}\textbf{LC: #1}}}
\else
\newcommand{\mmm}[1]{}
\newcommand{\jmh}[1]{}
\newcommand{\pjj}[1]{}
\newcommand{\lc}[1]{} 
\fi

\newcommand{\rfone}[1]{#1}
\newcommand{\rftwo}[1]{#1}

\definecolor{myblue}{RGB}{0,154,249}
\definecolor{mygreen}{RGB}{0,128,0}
\definecolor{myorange}{RGB}{226,111,70}
\definecolor{mymagenta}{RGB}{195,113,211}

\newcommand{\dd}{\mathrm{d}}
\newcommand{\pd}{\partial}

\renewcommand{\vec}[1]{\overline{#1}} 
\newcommand{\cov}[1]{\underline{#1}} 
\newcommand{\ee}{\vec e} 
\newcommand{\cee}{\cov e} 

\newcommand{\hh}{h}
\newcommand{\ff}{f}
\newcommand{\fdist}{\mathcal F}

\begin{document}

\title{Fully-implicit Particle-in-Cell model of a Magnetic Nozzle with electromagnetic power deposition}

\author[1]{Mario Merino}

\author[1]{Juan Martín-Hernández}

\author[1]{Pedro Jiménez-Jiménez}

\author[2]{Luis Chacón}

\affil[1]{Universidad Carlos III de Madrid, Department of Aerospace Engineering, 28911 Leganés (Madrid), Spain}  
\affil[2]{Theoretical Division, Los Alamos National Laboratory, New Mexico 87544, United States}  
 
\maketitle

\begin{abstract}
A fraction of the electromagnetic power used to generate and heat the plasma in helicon sources and electrodeless plasma thrusters can leak into the outer expansion region, interacting with the plasma in the magnetic nozzle and affecting the performance of the device.
This work analyzes the properties of the plasma in a convergent-divergent magnetic nozzle when right-hand polarized waves of varying amplitude propagate into it.
This is accomplished with a 1D3V fully-implicit, Vlasov-Darwin particle-in-cell model of the collisionless ion and electron plasma in a magnetic tube. The code exactly conserves charge locally and energy globally. It features a nonuniform grid and an enhanced substepping routine for the macroparticle trajectories. The requirement that the expansion be current-free is satisfied thanks to linear closed-loop controllers on the injection and downstream boundary conditions.
Wave heating increases the electron perpendicular temperature, especially in the vicinity of an electron cyclotron resonance surface, always present inside the magnetic nozzle of a helicon device. The energized electrons 
become anisotropic, and
drive a more pronounced potential drop and a higher ion acceleration than in the absence of waves, at the expense of the wave power. 
The computed moments of the ion and electron distributions reveal the dominant balance of the electron thermal terms, electrostatic terms, and ion inertial terms in the momentum and energy equations.
Wave heating helps populate otherwise-inaccessible regions of the electron phase space and modifies the doubly-trapped electron population found in the purely electrostatic case.
The overall efficiency of the magnetic nozzle increases mildly at low wave power, but drops quickly as it is increased beyond a threshold, as part of the energy is wasted in the form of parallel electron pressure and heat flux.
We further show that it is possible to obtain similar wave fields as in the kinetic model with a simple cold plasma linear dielectric model, albeit at the expense of self consistency and an empirically-fixed collisionality parameter.
Computational scaling estimates show that the fully-implicit PIC model \rfone{might} facilitate realistic simulations of plasma thrusters in one or more dimensions, which can be too costly with explicit codes. 
\end{abstract}

\section{Introduction}\label{sec:intro}

Electrodeless Plasma Thrusters (EPTs) are a class of electric propulsion devices that do not require hollow cathodes for ion beam neutralization \cite{ahed11s,mazo16a}. This brings in the potential of simplified power processing units, enhanced lifetimes, and the flexibility to use alternative propellants.
Among the various EPT designs, the Helicon Plasma Thruster (HPT) \cite{taka13b,nava18a,shin14a} and the Electron Cyclotron Resonance Thruster (ECRT) \cite{mill66,serc93,inch21b,desa23a} have gained the attention of the research community in the past decade.
A key feature of these thrusters is the utilization of electromagnetic waves to energize the charged particles
\cite{bath17a,taka19}, along with a downstream plasma expansion into vacuum guided by a magnetic nozzle (MN)
\cite{kuri70,ahed10f,taka12,meri16g,litt19}.
The major difference between the HPT and ECRT is their operating frequencies (in the MHz and low GHz ranges, respectively), and power coupling mechanisms. 

Understanding the physics of the discharge and, in particular, the wave-plasma interaction, remains a significant challenge for the design and optimization of these devices.
The wave-plasma interaction has been previously modeled using simplified linearized cold-plasma models in the frequency domain, sometimes coupled with fluid electron transport solvers \cite{tian18a,svil21a,jime23a,inch24b}. 
In HPTs, inductors (or antennae) around the cylindrical plasma source couple the electromagnetic power with the plasma, in which right-hand polarized waves of large amplitude propagate. Absorption can occur due to electron collisionality \cite{tian18a} and through resonant wave-particle interactions, although it is also known that parametric decay instabilities and other nonlinear phenomena can play a major role, at least in helicon devices \cite{pork72,ligh01,madd25a}.
In ECRTs, most of the power deposition on the plasma takes place
at the resonance surface for the right-hand polarized wave, located inside the discharge chamber. The absorbed power determines the electron temperature profiles, which are responsible for ionization and, to a large extent, the performance of the device. Recent work shows that the electron temperature peaks near the inner conductor in coaxial-type ECRTs, in agreement with experimental observations \cite{svil21a}.
Interestingly, this principal resonance is also present in HPTs, although for the typical frequencies and magnetic field strengths, it is located somewhere downstream, inside the MN. 
Cold-plasma wave models coupled with plasma transport models have shown that 
the wavefields in the HPT can easily propagate up to this surface, and their absorption can result in a localized increase of the electron temperature, a feature that was also found experimentally \cite{jime23a}.

While effective to understand leading-order effects, these simplified and linearized wave simulation approaches fail to capture nonlinear effects and kinetic phenomena, and often have free parameters that must be empirically selected. 
To overcome these limitations, self-consistent, fully nonlinear, kinetic-electromagnetic simulations are required. Due to the high computational cost, kinetic simulations are scarce, and most of the existing ones are only electrostatic, ignoring the wave-plasma interaction phenomena.
Relevant works include those of Martínez-Sánchez et al. \cite{mart15}, Sánchez-Arriaga et al. \cite{sanc18b,zhou21a}, Ramos et al. \cite{ramo18a} and Merino et al. \cite{meri21a}, which rely on different forms of Eulerian solvers of the Vlasov-Poisson system to solve the distribution function of ions and electrons in the MN of these devices. These studies show how the electron population in the plume is divided into free streaming electrons capable of balancing the ion current, and a majority of reflected and trapped electrons which follow different dynamics and cooling laws.
Also used in the simulation of these devices are electrostatic particle-in-cell (PIC) codes, which have been successfully applied to the simulation of plasma physics inside thrusters and their plumes \cite{hu17, carl18, cich22a, andr22, mari23a}. 
The treatment of open boundaries such as the one found downstream of the simulation domain merits special care, as the electron species is dominantly a confined one, and electrostatic reflections \textit{beyond} the end of the domain are essential to determine the electron population consistently inside it. In addition, since the population of reflected electrons is not exactly known a priori, the injection of macroparticles upstream must be adapted to account for it as it forms. Boundary conditions that take these facts into account and selectively reflects part of the escaping electrons back into the domain according to an energy threshold were successfully tested in \cite{li19a}.
The work of Andrews et al. \cite{andr22} later refined the downstream boundary condition on the electrostatic potential, including first-order electric monopole corrections. Jim\'enez et al. \cite{jime24a} formulated a self-consistent boundary condition for 1D MNs that used Ampère's equation to predict the half-time electric field based on the electric current leaving the domain.

However, none of these studies examines the electromagnetic wave-plasma interaction.
To our knowledge, the only existing example of a kinetic-electromagnetic study of EPTs is the recent work by Porto et al. \cite{port23}. 
That study employs 1D explicit particle-in-cell/Maxwell simulations with a Constrained Interpolation Profile (CIP) scheme to model transverse fields propagating through a plasma in the MN of a coaxial ECRT. Their results show anisotropic plasma heating inside the thruster and a trapped population of electrons with higher perpendicular energy in the plume. The CIP method alleviates the CFL condition for fast wave propagating modes, but it does not eliminate the need to solve quite short time and length scales. 

All the PIC studies listed so far  (with the exception of \cite{jime24a}) employ a time-explicit leapfrog scheme \cite{BIRD91}. Time-explicit PIC codes suffer from stringent timestep and cell-size constraints, requiring resolution of the electron plasma frequency and the Debye length for stability. As such, their computational demands are huge and their application is limited to small domains or to problems with modified vacuum permittivity, $\varepsilon_0$. 

In contrast, time-implicit PIC algorithms \cite{chen11,lape17,erem22,jime24a}
can  step over the small temporal and spatial scales if their resolution is not necessary to understand the physics of the scales of interest, and thus
are a powerful tool to overcome the limitations of time-explicit PIC codes and conduct realistic simulations with reasonable computational resources. 
This is owing to the fact that the implicit time-stepping disperses the fast plasma frequency oscillations under large time steps. Similarly, it is possible to formulate the model to be able to use cell sizes larger than the Debye length without incurring finite grid instabilities \cite{huangFiniteGridInstability2016,barn21}.

Time-implicit PIC codes have been developed and applied in
various research fields, including nuclear fusion \cite{chen23b} and material processing \cite{erem22}. Their application to electric propulsion studies has been relatively unexplored until recently \cite{tacc23,jime24a}. These algorithms were initially developed for electrostatic scenarios \cite{chen11}, but recent studies in one and multiple dimensions \cite{chen14, chen15a} have shown that incorporating the electromagnetic fields through the Darwin approximation, which excludes light-speed traveling modes  \cite{hewe85, krau07}, enables efficient and accurate kinetic electromagnetic simulations over long time periods.
An important feature of many of the existing time-implicit PIC algorithms is that they can easily be constructed to conserve charge locally and energy globally (and in some special cases, locally as well \cite{chacon2025local}) to machine precision, which is desirable for long-term simulation.
In contrast, most time-explicit PIC codes, while momentum-conserving, are not energy-conserving.

This work presents a collisionless 1D-3V time-implicit electromagnetic code and applies it to the study of a MN, into which a right-hand polarized wave propagates. Downstream, as the magnetic field strength decreases, the electron-cyclotron resonance conditions are eventually met, and a strong coupling between the wave and the plasma takes place.
This is the situation in the plume of an HPT that leaks RF power into the MN, referred above. 
\rfone{Moreover, the problem under analysis could tangentially shed some light on
some of the physics of the wave-plasma interaction inside an ECRT,
which has a similar diverging magnetic field topology and resonance, albeit ignoring the otherwise-important collisions in the denser source plasma and its interactions with the thruster walls.}
The goal of the study is to understand the effects of the propagation and absorption of the wave on the expanding magnetized plasma, and in particular, on the electrons. The first fluid moments of the electron distribution and the balances in the momentum and energy equations are 
analyzed. The wavefields are compared with the cold-plasma dielectric tensor theory, showing that a good approximation of the self-consistent fields is possible once an effective damping parameter in that model is tuned.
This is a key finding that exemplifies how kinetic simulations can be used to inform and fit faster, simpler surrogate models. 
The direct analysis of the electron distribution function  reveals how the anisotropy in the MN is altered by the wave power, how a non-uniformity in the gyrophase exists near the resonance surface, and how the wave can modify the balance between free, reflected, and trapped electrons. The hotter electrons (at the expense of the wave power) drive a larger acceleration of the ion beam, but the performance of the MN finds an optimum at relatively mild wave power levels.

As a secondary objective of this work, the findings and lessons learned help outline the path toward the fully-implicit, kinetic-electromagnetic simulation of full EPTs in higher dimensions. 
We briefly extend the discussion in \cite{jime24a} on how the time-implicit electromagnetic PIC approach
can achieve significant computational savings while maintaining accuracy and conservation properties.
Indeed, our approach builds on top of our recent fully-implicit, kinetic-electrostatic 1D investigation of magnetized plasma expansions \cite{jime24a}, extending the previous simulation framework with an electromagnetic Darwin model. Rather than integrating in the magnetic moment of the particles and their axial velocity, we base our formulation on Cartesian velocity components.
New boundary conditions for the electromagnetic fields are  introduced, as well as refined closed-loop controllers to verify the fulfillment of global conditions in the MN. 

The remainder of the paper is organized as follows. Section
\ref{sec:problemstatement} presents the physical problem under study in more detail. Sections
\ref{sec:model} and \ref{sec:discretization} respectively introduce
the collisionless, 1D-3V, fully-implicit, electromagnetic PIC model and its numerical discretization.
Section \ref{sec:results} presents and discusses simulation results for various wave power levels entering the MN.
The convergence and scaling of the model upon varying the numerical parameters is briefly visited in Section \ref{sec:verification}.
Finally, conclusions are gathered in Section \ref{sec:conclu}. 
Appendices detail the differential geometry of the computational coordinates used in this work (App. \ref{sec:diffgeometry}) and the key properties of the B-spline shape functions used (App. \ref{sec:shapefunctions}).

\section{Problem overview}\label{sec:problemstatement}

\begin{figure}
    \centering
    \includegraphics[width=0.9\linewidth]{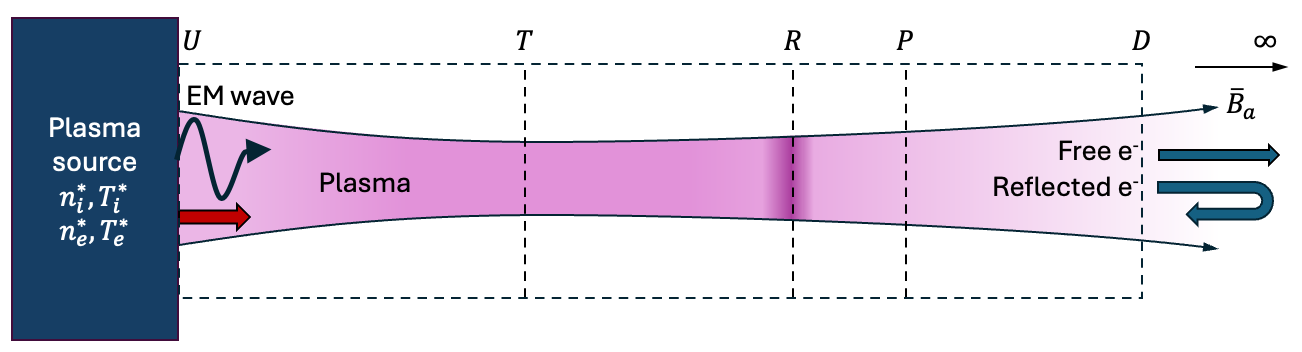}
    \caption{Sketch of the problem. An upstream plasma source with a Maxwellian distribution of ions and electrons (given by $n_i^*,T_i^*$ and $n_e^*,T_e^*$) is injected into a convergent-divergent magnetic tube at an upstream section $U$. At the same location, an electromagnetic wave of frequency $\omega$ leaks into the domain. The applied magnetic field $\vec B_a$ reaches a maximum strength at the magnetic throat $T$, where the magnetic tube section is minimal. Electron-cyclotron resonance conditions $\omega=\omega_{ce}$ are met at station $R$. Station $P$ indicates the location where the wave matches the plasma frequency $\omega=\omega_{pe}$. The end of the simulation domain is $D$; the plasma expansion continues to infinity beyond, $\infty$.}
    \label{fig:domainsketch}
\end{figure}

A converging-diverging, axisymmetric MN, created by an applied magnetic field $\vec B_a$, is connected to an upstream plasma reservoir with electrons ($e$) and singly-charged ions ($i$) with prescribed properties. We denote the direction of the axis of the MN with coordinate~$z$.
Figure \ref{fig:domainsketch} sketches the main elements of the problem considered in this work.

In the MN, the warm, expanding electrons set up a downstream-pointing electrostatic field $\vec E_s= -\nabla \phi$, which confines most of them and simultaneously accelerates the ions downstream. The electrostatic potential $\phi$, which must be found as part of the solution, is such that it ensures that the net currents of ions and electrons are equal, giving as a result a globally current-free plasma expansion.

In an efficient MN operating in space, the collisional mean free path is large away from the throat, and so particle collisions play a minor role~\cite{meri16g}.
As the plasma expands collisionlessly, guided by the MN, the part of the distribution functions $\fdist_e$, $\fdist_i$ with $v_z\geq 0$ are assumed to be known upstream, and act as boundary conditions for the MN problem. 
On the other hand, the part of the distribution functions with $v_z<0$ at this location is unknown and depends on the existence of reflected particles in the MN traveling back toward the source, if any, which are part of the solution.
Indeed, in a stationary electrostatic setting where the mechanical energy of the individual particles is conserved, the combined effect of the magnetic mirror created by $\vec B_a$ and the electrostatic mirror created by $\phi$ subdivides the electrons into three distinct populations according to their energy and magnetic moment \cite{mart15, sanc18b, ramo18a, ahed20a, meri21a}:
(i) \textit{free} electrons, with a mechanical energy high enough to overcome the potential barrier and reach infinity downstream;
(ii) \textit{reflected} electrons, with lower energy and whose trajectories connect with the upstream reservoir; and
(iii) \textit{trapped} electrons, with low energy but high magnetic moment, whose trajectories connect neither with the upstream reservoir nor with infinity.
In contrast, in the diverging MN ions are accelerated downstream both by the magnetic mirror and the electrostatic field, and so all ions are free, with the exception of a small amount of ions that can be reflected in the convergent side, before reaching the throat.

A forward-traveling, right-hand  polarized (RHP) electromagnetic wave of frequency $\omega$ and known magnitude propagates from the reservoir into the MN, as the result of an incomplete absorption of the applied RF power in the plasma source. This propagation goes on while $\omega_{ce} = eB_a/m_e > \omega$ (i.e., the local electron cyclotron frequency is greater than the wave frequency); however, as the magnetic field strength diminishes, the electron-cyclotron resonance (ECR) condition $B_a = m_e\omega/e$ is ultimately met at a downstream section, and beyond that point, the wave becomes evanescent. Propagation of waves formally resumes beyond the section at which the decreasing plasma frequency, $\omega_{pe}=\sqrt{n_e e^2/(\varepsilon_0 m_e)}$, matches the wave frequency; the location where this critical density occurs depends on the profile of $n_e$, and thus it is not known a priori. 

The wave fields 
$(\vec E_w, \vec B_w)$
and the plasma particles (especially the electrons) interact in the MN, exchanging energy, and therefore the wave offers a mechanism to transfer electrons among the free, reflected and trapped regions of phase space. Most of the wave power is expected to be absorbed at or near the ECR surface, where neither the energy nor the magnetic moment of electrons is conserved, resulting in the anisotropic heating of the electrons.

\section{Model}\label{sec:model}

The region of interest of the present model is a thin magnetic tube of radius comparable to the largest Larmor radius in the problem $\ell$ around the axis of the MN, which coincides with the $z$ axis. In this tube, $x$, $y$ and $r=\sqrt{x^2+y^2}$ are of order $\ell$.
The tube starts at an upstream section in the convergent side of the magnetic field (subscript $U$ in figure \ref{fig:domainsketch}), passes through the magnetic throat where $B_a$ is maximum (subscript $T$) and expands to infinity (subscript $\infty$). However, our model shall only solve the plasma properties in a finite domain, truncating the expansion at a downstream boundary (subscript $D$). In between, the resonance condition (subscript $R$) and the critical density condition (subscript $P$) are found. 
We place our origin of coordinates at the throat $T$, where $z=0$.
The following orderings of length scales are assumed to hold everywhere in this tube:
\begin{align}
\ell  \sim \varepsilon L \sim \varepsilon^2 \lambda_{c},  
\end{align}
where $L$ is the fundamental, characteristic gradient length, which we take to be $L = (\pd_z \ln B_a )^{-1}$; $\lambda_{c}$ is the smallest collisional mean free path; and $\varepsilon$ is a small expansion parameter. 
Under these assumptions, the plasma expansion can be treated as \textit{well-magnetized} and \textit{collisionless}. 
Furthermore, owing to the axisymmetry of the problem, any radially-even profile $Q$ (e.g. the electron density $n_e$ or temperature $T_e$) is uniform in $r$ to order $O(\varepsilon^2)$ within such a tube, i.e. $Q(z,r) = Q(z,0)+ O(\varepsilon^2) (r/\ell)^2$.
Notably, we do not prescribe any ordering for the Debye length $\lambda_{De}$ nor impose plasma quasineutrality in the bulk; however, the results in Section \ref{sec:results} confirm that the plasma happens to be essentially quasineutral everywhere.

Since the magnetic moment $\mu$
of each particle is not conserved in this problem because of the interaction with the wave, and moreover, the distributions may develop non-uniformities in the gyrophase $\alpha$, we choose to work in Cartesian position space coordinates $(x,y,z)$ and Cartesian velocity space coordinates ($v^z,v^x,v^y$) rather than in $(z,r)$ and $(v^z,\tilde\mu)$ as in \cite{jime24a}. 
Moreover, we define the orthogonal computational coordinates 
($\xi, \eta, \zeta$), which within the small magnetic tube map to the physical Cartesian coordinates as:
\begin{align}
x &= \hh \xi,
&
y &= \hh \eta,
&
z &= \int_0^\zeta \ff\dd \zeta -\frac{\hh\hh'}{2\ff}(\xi^2 + \eta^2).
\label{eq:changeofcoordinates}
\end{align}
where the Lamé factors $\hh,\ff$ are such that $\hh(\zeta) = \sqrt{B_{aT}/B_a(\zeta)}$ scales with the radius of the magnetic field tube, and $\ff(\zeta)$ can be chosen to differ from $1$ to facilitate the introduction of non-uniform grids in Section \ref{sec:discretization}. These coordinate definitions are exact at the axis but we accept that they have an error $O(\varepsilon^3)$ away from it to keep the coordinates $\xi,\eta,\zeta$ strictly orthogonal everywhere.
The differential geometry induced by this coordinate transformation is detailed in Appendix \ref{sec:diffgeometry}, including the  definition of the vector basis $\{\ee_\xi,\ee_\eta,\ee_\zeta\}$ and the covector basis $\{\cee^\xi, \cee^\eta, \cee^\zeta\}$.
The appendix also reduces the expression of the gradient, divergence, curl, and Laplacian differential operators used in the rest of this work.

Since $\zeta$ lines are $\vec B_a$ lines, the applied magnetic field $\vec B_a$ has only one non-zero component, $B^\zeta$, in the local basis. Consequently, the Cartesian components, with error $O(\varepsilon^2)$ away from the axis, are:
\begin{align}
B_a^x &=\xi \hh' B_a^\zeta;
&
B_a^y &= \eta \hh' B_a^\zeta;
&
B_a^z &= \ff B^\zeta. 
\label{eq:B_a_xyz_components}
\end{align}
In the following, and to lighten our formulation, we omit the error order in $\varepsilon$ in each expression, as it can be inferred from the information in Appendix \ref{sec:diffgeometry}.

\subsection{Electromagnetic field equations}\label{sec:fields}

The total electric and magnetic fields can be split into their various contributions:
\begin{align}
\vec E &= \vec E_s + \vec E_w; 
&
\vec B &= \vec B_a + \vec B_w.
\end{align}
The vectors $\vec {E_s}$ (electrostatic field), $\vec E_w$, and $\vec B_w$ (wave fields) admit a representation based on the electrostatic and magnetic potentials $\phi$ and $\vec A$ (in the Coulomb gauge, such that $\nabla \cdot \vec A=0$):
\begin{align}
\vec E_s &= -\nabla \phi; 
&
\quad \vec E_w &= - \pd_t \vec A;
&
\quad \vec B_w &= \nabla\times \vec A.
\label{eq:fieldsdef}
\end{align}

To derive the equations for the potentials, 
consider first Poisson's equation and the complete Ampère's law:
\begin{align}
-\varepsilon_0 \nabla^2 \phi &= \rho,
\label{eq:Poisson}
\\
\varepsilon_0\mu_0\left(\pd_t\nabla\phi  + \pd^2_t\vec A\right) + 
\nabla\times(\nabla\times \vec A)
&= \mu_0\vec{\jmath}.
\label{eq:Ampere}
\end{align}
where $\rho$ is the charge density  and $\vec{\jmath}$ is the charge current density, defined in Section \ref{sec:kineticeqs}.

The full equation \eqref{eq:Ampere}, which is manifestly hyperbolic for $\vec A$, introduces the time scale associated with the speed of light, $c=1/\sqrt{\varepsilon_0\mu_0}$, much faster than any characteristic time scale of interest in the plasma problem.
This sets an additional burden to the computational solution of the fields, which in many scenarios (like the present one) is not warranted.
To ameliorate this, we apply Darwin's approximation, which entails dropping the solenoidal $\pd^2_t\vec A$ term (but not $\pd_t\nabla\phi$) in the displacement current in equation \eqref{eq:Ampere} as in \cite{chen14b,chen15a}. This simplification is of $O(v/c)^2$ \cite{krau07}, and commonly used in the study of low-frequency electromagnetic fields in dense plasmas due to its computational advantages, since in this non-radiative regime the electric current $\vec\jmath$ dominates over the displacement current \cite{HOCK88, krau07}.
Indeed, for planar waves, the consequence of this approximation is the shifting of the squared dimensionless refraction index by $1$ unit,
which is a negligible modification inside a plasma away from cutoffs. A related, second effect of Darwin's simplification is that vacuum light waves are removed from the problem. Thirdly, and relevant from a numerical standpoint, dropping this term changes the nature of the resulting electromagnetic equations from hyperbolic to elliptic in the potentials. 

To eliminate the need of computing both $\rho$ and $\vec\jmath$ during the solution of the problem, we 
follow \cite{chen14b,chen15a} and take the divergence of equation \eqref{eq:Ampere}, obtaining the evolution equation for $\phi$:
\begin{align}
\varepsilon_0\pd_t\nabla^2\phi =   \nabla \cdot \vec \jmath,
\label{eq:phi's equation}
\end{align}
Observe that this equation is also obtained
(and therefore is equivalent to) taking the time derivative of equation \eqref{eq:Poisson} and invoking the conservation of electric charge, 
\begin{align}
 \pd_t \rho + \nabla\cdot\vec\jmath = 0.
 \label{eq:continuity}
\end{align}
Incidentally, the reason why the $\pd_t\nabla\phi$ term cannot be simply dropped like the $\pd^2_t\vec A$ term in equation \eqref{eq:Ampere} is that it would result in the violation of the conservation of charge (or, equivalently, an incompatibility between Ampère's and Poisson's equations).

We next establish the functional structure of the potentials $\phi$ and $\vec A$ in the neighborhood of the axis as described
in Appendix \ref{sec:diffgeometry}, by prescribing
$\phi=\phi(\zeta)$ and $\vec A = \vec A (\zeta)$ (i.e., we drop the higher-order radial dependency at the axis).
This fixes the divergence rate of the streamlines of $\vec E_s$ and the wave power to match the divergence rate of the applied magnetic field streamlines, which is a reasonable choice in view of existing fully-magnetized MN models \cite{meri16a}.
Furthermore, the Coulomb gauge demands $\vec A$ to be solenoidal, $\nabla\cdot\vec A = 0$. This condition is satisfied if $A^\xi=A^\xi(\zeta)$,
$A^\eta=A^\eta(\zeta)$, and $A^\zeta=C/(\hh^2\ff)$ with $C=C(t)$.
Without loss of generality in the solution we take $A^\zeta=0$, which cleanly separates the contributions of $\phi$ and $\vec A$ to the electric field, with  $\vec E_s 
= -\nabla\phi$ and $\vec E_w 
= -\pd_t \vec A$.

With these provisions, and writing the vector result back in Cartesian components, the equations for $\phi$, $A_x$, and $A_y$ reduce at the axis to:
\begin{align}  
\pd_t\pd_\zeta\left ( \frac{\hh^2}{\ff} \pd_\zeta \phi \right) &= \frac{1}{\varepsilon_0}\pd_\zeta \left(\hh^2 j^z\right),
\label{eq:phiz}
\\
\pd_\zeta\left [\frac{1}{\ff} \pd_\zeta\left(\hh A^x\right) \right] &= -\mu_0 \hh \ff j^x, 
\label{eq:Ax}
\\
\pd_\zeta\left [\frac{1}{\ff} \pd_\zeta\left(\hh A^y\right) \right] &= -\mu_0 \hh \ff j^y.
\label{eq:Ay}
\end{align}
Note that equation \eqref{eq:phiz} can be integrated once in $\zeta$ to yield the (time derivative of the) Gauss integral law for the electric field, but care must be taken with solvability constraints and boundary conditions \cite{jime24a}.

The boundary conditions for the electrostatic potential $\phi$ are as follows. The upstream potential at $U$ is fixed to a reference value, $\phi=0$ (Dirichlet condition). 
In a small departure from our previous work \cite{jime24a}, the downstream boundary condition at $D$, $\phi=\phi_D$, is determined as part of the solution with the condition that the end of the domain must remain quasineutral. This was found to result in smoother transients than our previous approach, which fixed $E^z_s$ at the downstream boundary following an integral Gauss condition. 
In either case, $E^z_s\simeq 0$ there, once steady state is reached.
Note that $\phi_D$ differs from $\phi_\infty$ at infinity, far beyond the domain, which needs to be taken into account to correctly determine the population of reflected electrons as indicated in Section \ref{sec:kineticeqs}.

Regarding the boundary conditions for $\vec{A}$, time-varying Dirichlet conditions for a right-hand polarized wave at a  frequency $\omega$ at the upstream boundary are imposed. 
This means that, on the axis at that location (subscript $U$):
\begin{align}
A^x - i A^y &= A_U \exp(-i \omega t).
\end{align}
To relate the amplitude $A_U$ to the input wave power per unit area, $P_{wU}$, we first write the Poynting vector as $\vec {\mathcal S}= \vec E_w \times \vec B_w/\mu_0 = -\pd_t \vec A\times(\nabla\times \vec A)/\mu_0$.
Accepting that the input wave locally has a single wavevector $k$ in the neighborhood of the upstream section $U$, after some algebra we approximate:
\begin{align}
\mathcal S_U \simeq
\frac{\omega k A^2_U}{\mu_0} 
\label{eq:S0estimate}
\end{align}
The right-hand polarized wave dispersion relation is used to compute $k=k_R$ at $U$, from
\begin{align}
k_R^
2(\zeta) = \frac{\omega^2}{c^2}\left(1 - \frac{\omega_{pe}^2(\zeta)}{\omega[\omega-\omega_{ce}(\zeta)]}\right).
\label{eq:kR}
\end{align} 
The first term in the parenthesis can be neglected with respect to the second term; in fact, this approximation is automatic in the Darwin model.

Finally, assuming that negligible wave power traverses the resonance to reach the downstream boundary after a long evanescent region, we simply set $A^x=A^y=0$ at the axis there.
\rfone{This simplified boundary condition is justified a posteriori by (1) the smallness of the $\vec A$ field in the downstream part of the domain
(in part due to the efficient absorption of power at the resonance R), which means that any reflection at the boundary has a negligible effect on the simulation as shown in Section \ref{sec:results}, and (2) the fact that extending the length of the domain does not change the solution of $\vec A$, as observed in Section \ref{sec:verification}.}
We note that an improvement over these boundary conditions would be to consider waveport boundary conditions at both ends, allowing for any reflected and transmitted power to escape from the domain without inducing spurious reflections at the boundaries. This may be warranted for other wave-plasma interaction problems;
nevertheless, our simulation results prove the validity of our assumption in the present problem under study. 
 
\subsection{Electron and ion kinetic equations}\label{sec:kineticeqs}

While position space is conveniently parametrized in the computational coordinates $\{\xi,\eta,\zeta\}$, it is preferable to keep Cartesian coordinates for velocity space
\cite{chacon2013charge,chac16}, $\{v^x,v^y,v^z\}$. This avoids introducing fictitious coordinate or curvature forces, which appear in the expression of the acceleration when written in terms of 
$\{\ddot\xi,\ddot\eta,\ddot\zeta\}$.
Moreover, this is a convenient choice in anticipation of a future extension of our model to higher dimensions, where Cartesian velocity components are a natural option. 

The \textit{gyrocenter} and \textit{particle} velocity distribution functions of ions and electrons, $\fdist_{g\varsigma}$ and $\fdist_\varsigma$, with $\varsigma=i,e$, are even functions of $\xi$ 
and $\eta$
at the axis.
Therefore, in the neighborhood of the axis $\fdist_\varsigma = \fdist_{g\varsigma}$  with error $O(\varepsilon^2)$. Since neither particles nor gyrocenters in this tube drift far from the axis, and since all dynamic variables of interest are independent of $\xi,\eta$ to said order, we can ignore the coordinates $\xi,\eta$ of particles and track only their coordinate $\zeta$.
Accepting that all magnetic lines around the axis expand likewise, this allows us to write simply $\fdist_\varsigma=\fdist_\varsigma(\zeta,v^x,v^y,v^z,t)$ in the tube, with an error $O(\varepsilon^2)$.  
 
Again with error $O(\varepsilon^2)$, a charged particle whose gyrocenter is located on the axis has $\xi = - m_\varsigma v^y/(\hh q_\varsigma B_a^z)$ and $\eta = m_\varsigma v^x/(\hh q_\varsigma B_a^z)$  where $q_\varsigma=+e,-e$ for $\varsigma=i,e$ respectively.
Using the expression of the Cartesian components of $\vec B_a$ in equation
\eqref{eq:B_a_xyz_components}, 
the acceleration experienced by this particle
is 
\begin{align}
a^x &= \frac{q_\varsigma}{m_\varsigma} \left( E_{w}^x + v^yB_a^z
-v^z B_w^y \right)
- \frac{  v^xv^z}{\hh \ff} \hh',
\nonumber
\\
a^y &= \frac{q_\varsigma}{m_\varsigma} \left( E_{w}^y - v^xB_a^z
+ v^z B_w^x\right) 
- \frac{  v^yv^z}{\hh \ff} \hh',
\label{eq:accelerations}
\\
a^z &= \frac{q_\varsigma}{m_\varsigma} \left(E_{s}^z 
+ v^x B_w^y - v^yB_w^x\right)
+ \frac{  (v^x)^2+ (v^y)^2}{\hh \ff} \hh'.
\nonumber
\end{align}
where the last term in each equation is consequence of the magnetic mirror effect.
In general, the effect of the magnetic field of the wave, $B_w$ on the acceleration of the particles, is small with respect to that of the applied field, $B_a$, but we keep it here for completeness.
Within our order of accuracy, the expression of the acceleration for particles whose gyrocenter does not lie exactly on the axis is the same as in equations \eqref{eq:accelerations}.

With all these prescriptions, the evolution of $\fdist_\varsigma$ in the collisionless limit is given by the following form of Vlasov's equation,  
\begin{align}
\frac{\dd \fdist_\varsigma}{\dd t} =  \pd_{t} \fdist_\varsigma
+ \dot\zeta \pd_{\zeta} \fdist_\varsigma
+a^x \pd_{v^x} \fdist_\varsigma 
+a^y \pd_{v^y} \fdist_\varsigma 
+a^z \pd_{v^z} \fdist_\varsigma  
= 0.
\label{eq:vlasov}
\end{align}
for $\varsigma=e,i$, where $\dot\zeta = v^\zeta = v^z/\ff$ at the axis.

The integral moments of $\fdist_\varsigma$ over velocity space give
the density $n_\varsigma$, bulk velocity $u^i_\varsigma$, parallel and perpendicular pressure $p^\parallel_\varsigma$ and $p^\perp_\varsigma$, axial flux of parallel and perpendicular heat $q^\parallel_\varsigma$ and $q^\perp_\varsigma$, etc. At the axis we have, in particular:
\begin{align}
n_\varsigma &= \int \fdist_\varsigma \dd v^x \wedge \dd v^y \wedge \dd v^z,
\\
n_\varsigma u^i_\varsigma &= \int v^i \fdist_\varsigma \dd v^x \wedge \dd v^y \wedge \dd v^z,
\\
p^\parallel_\varsigma  &= m_\varsigma\int (c^z_\varsigma)^2 \fdist_\varsigma \dd v^x \wedge \dd v^y \wedge \dd v^z,
\\
p^\perp_\varsigma  &= \frac{m_\varsigma}{2}\int [(c^x_\varsigma)^2 + (c^y_\varsigma)^2] \fdist_\varsigma \dd v^x \wedge \dd v^y \wedge \dd v^z,
\\
q^\parallel_\varsigma  &= \frac{1}{2}m_\varsigma\int (c^z_\varsigma)^3 \fdist_\varsigma \dd v^x \wedge \dd v^y \wedge \dd v^z,
\\
q^\perp_\varsigma  &= \frac{1}{2}m_\varsigma\int [(c^x_\varsigma)^2 +(c^y_\varsigma)^2 ] c^z_\varsigma \fdist_\varsigma \dd v^x \wedge \dd v^y \wedge \dd v^z,
\end{align}
where $\vec c_\varsigma = \vec v - \vec u_\varsigma$
and $\varsigma=e,i$. Temperatures are defined as $T^i_\varsigma = p^i_\varsigma / n_\varsigma$. Other moments can be defined likewise. The charge density and current density in equations \eqref{eq:Poisson} and \eqref{eq:Ampere} are defined as 
$\rho = e(n_i-n_e)$ and $\vec j = e(n_i\vec u_i - n_e\vec u_e)$.

In terms of boundary conditions, the forward distribution function of ions and electrons at the upstream boundary, $\fdist_{iU}^+$ and $\fdist_{eU}^+$ must be supplied (i.e., for particles with $v^z>0$). In the present work, 
we fix the ion $\fdist_{iU}^+$ to be a half Maxwellian with amplitude $n_i^*$
and temperature $T_{i}^*$.
Note that, since most ions are free (with the exception of a few of them that are reflected back before the throat by the magnetic mirror effect), these values are related with the ion density and temperature upstream. In particular,  $n_{iU} \gtrsim n_{i}^*/2$.
The electron
$\fdist_{eU}^+$ is also modeled as a half Maxwellian, of amplitude $n_e^*$ and temperature $T_e^*$. Note also that, since a large fraction of the electrons is reflected back to the reservoir by the electrostatic potential $\phi$, the value of the electron density and temperature upstream, $n_{eU}$ and $T_{eU}$, are close to $n_e^*$ and $T_e^*$. 
The backward part of the distribution function, $\fdist_{eU}^-$,
depends on the potential far downstream $\phi_\infty$ which is unknown a priori.
This electron population also affects the electron density everywhere, 
which means that to keep the ion and electron density balanced upstream, $n_e^*$ needs to be computed as well as part of the solution. 
 
\section{Discretization and implementation}\label{sec:discretization}

Next, we describe how the Vlasov-Darwin system of equations is discretized and integrated in time. We begin with the discretization of the fields onto the grid; we then discuss how the macroparticle orbits are integrated. The computation of the sources on the Ampère equation (i.e., the charge current density) by weighting the macroparticles is described next. We follow with the treatment of boundary conditions and the associated closed-loop controls to satisfy the integral conditions.
Lastly, we briefly summarize the fully-implicit time integration scheme.
The resulting model conserves charge locally and energy globally (see \cite{chen14b,jime24a} for more details).

\subsection{Gridded functions}\label{sec:griddedfunctions}

We discretize the simulation domain into $N_g$ cells and $N_g+1$
nodes. Nodes correspond to integer positions; the upstream boundary corresponds to node $i=0$, while the downstream boundary corresponds to node $i=N_g$. Cells correspond to half-integer positions; first cell is $i=1/2$, and last cell $i=N_g-1/2$.
The present implementation of the model can work with analytical spatial grid maps $\ff(\zeta)$, in a similar way as in \cite{chac16, jime24a}. This enables using non-uniform grids to locally refine regions of higher field gradients (e.g. the upstream region) and maintaining a good balance of macroparticles per cell that takes into account the plasma expansion.

\begin{figure}
\centering
\includegraphics[width = 0.8\textwidth]{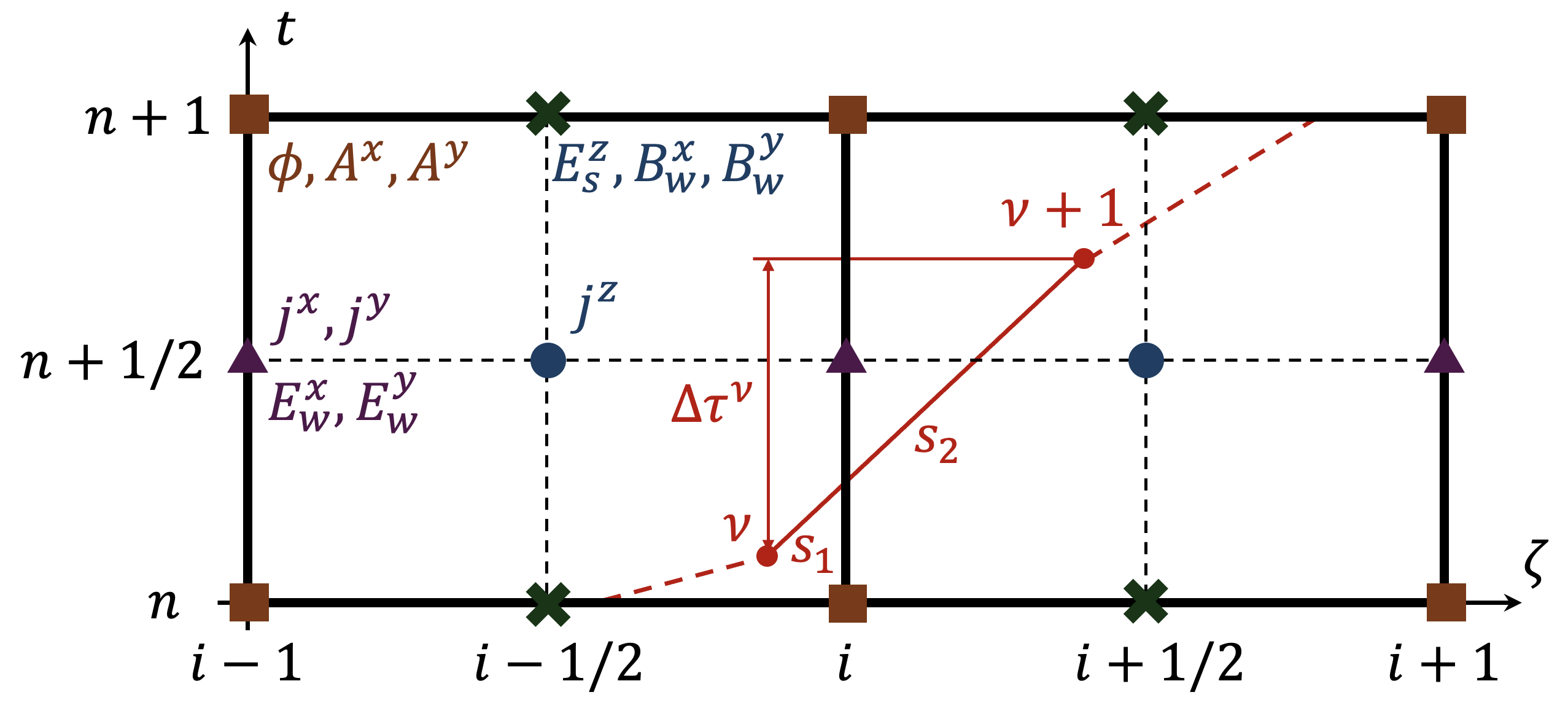}
\caption{Sketch of two cells in the computational (or logical) spatiotemporal grid.
Index $i$ is used to denote the spatial location, and index $n$ the time instant.
At the nodes, the potentials $\phi,A^x, A^y$ are computed
at integer times (dark red squares), and the current densities $j^x, j^y$ and the fields $ E^x_w, E^y_w$ at half-integer times (purple triangles).
At cell centers, fields $E^z_s$, $B_w^x$, $B_w^y$ are taken at integer times (green crosses), and the current density $j^z$ at half-integer times (blue dots).
A sample macroparticle trajectory is displayed in red. The trajectory is composed of multiple substeps, and a substep $\nu$ with time duration $\Delta \tau^\nu$ is highlighted.
Substeps can span across several cells, resulting in two or more segments $s$.}
\label{fig:spacetimesketch}
\end{figure}

A crucial choice in any particle-in-cell code is at what
grid location/time every  
grid function, such as the electromagnetic fields $\vec E_s, \vec E_w$, potentials $\phi, \vec A$, and sources $\vec \jmath$, is discretized and computed.
A good discretization ensures the correct space and time centering of finite differences in all the definitions and equations. 
As in earlier studies, in the present work  $\phi$, $A^x$, $A^y$, $E^x_w$, $E^y_w$, $j^x$, $j^y$ are taken at the grid nodes (integer positions, $\zeta=i$), and $E^z_s$, $j^z$, $B_w^x$, $B_w^y$ at the cell centers (half-integer positions, $\zeta=i+1/2$).
While $\phi$,  $A^x$, $A^y$, $E^z_s$, $B_w^x$, $B_w^y$ are computed at integer time instants ($t=n\Delta t$), the charge current densities  $j^x$, $j^y$,  $j^z$ and the wave fields $E_w^x$ and $E_w^y$ are computed at half-integer time instants ($t=(n+1/2)\Delta t$). 
Figure \ref{fig:spacetimesketch} sketches the spatiotemporal location of our discrete variables.
In this manner, using $i$ for the spatial grid index and $n$ for the time index, the field definitions in equation \eqref{eq:fieldsdef} read:
\begin{align}  
E_{s,i+1/2}^{z,n} &= - \frac{\phi^n_{i+1}-\phi^n_{i}}{\ff_{i+1/2}},
\\
E_{w,i}^{x,n+1/2} &= -\frac{1}{\Delta t}\left( A_{i}^{x,n+1} - A_{i}^{x,n} \right),
&
E_{w,i}^{y,n+1/2} &= -\frac{1}{\Delta t}\left( A_{i}^{y,n+1} - A_{i}^{y,n} \right),
\\
B_{w,i+1/2}^{x,n} &= -\frac{\hh_{i+1}A_{i+1}^{y,n} - \hh_{i}A_{i}^{y,n}}{\hh_{i+1/2}\ff_{i+1/2}},
&
B_{w,i+1/2}^{y,n} &= \frac{\hh_{i+1}A_{i+1}^{x,n} - \hh_{i}A_{i}^{x,n}}{\hh_{i+1/2}\ff_{i+1/2}},
\end{align}
where we have used $\Delta\zeta = 1$.
The $z$ component of Ampère's equation \eqref{eq:phiz} discretizes as follows:
\begin{align}  
\hh_{i+1/2}^2 (E_{s,i+1/2}^{z,n+1}-E_{s,i+1/2}^{z,n}) -\hh_{i-1/2}^2 (E_{s,i-1/2}^{z,n+1}-E_{s,i-1/2}^{z,n}) &= 
\nonumber
\\
-\frac{\Delta t}{\varepsilon_0}  \left(\hh_{i+1/2}^2 j_{i+1/2}^{z,n+1/2}-\hh_{i-1/2}^2j_{i-1/2}^{z,n+1/2}\right).
\label{eq:phiz_discrete}
\end{align}
Note that the chosen positioning of the variables on the grid offers indeed a naturally well-centered discrete version of all equations listed so far. 

This leaves us to discuss the remaining two components of the Ampère equation, \eqref{eq:Ax} and \eqref{eq:Ay}.
At this point, it becomes evident that a correct centering of the finite differences for these equations is not possible. Indeed, centering in this case would require defining $A^x$, $A^y$ at half-integer times (where the currents are defined), and consequently, taking $E^x_w$, $E_w^y$ at integer times. 
However, doing so means that computing the fields $E^x_w$, $E_w^y$ at time $t=n\Delta t$ requires knowledge of $A^x$, $A^y$ at $t=(n+1/2)\Delta t$, which only becomes available in the subsequent timestep. This would break the time-marching approach needed to compute the solution. 
The underlying reason for this is the elliptic nature of the Ampère equation in the Darwin approximation, which does not couple the components of $\vec A$ across time.  
In contrast, the full Maxwell system that retains the $\pd_t^2 \vec A$ term in equation \eqref{eq:Ampere} is hyperbolic on $\vec A$. Indeed, in the full Maxwell, said placement of the potentials and the fields becomes natural, and a time-marching solution is possible.

To overcome this difficulty, we proceed as in \cite{chen14b} and interpolate the $A^x$ and $A^y$ potentials at half-integer times. We do this by defining the averages:
\begin{align}
A_{i+1}^{x,n+1/2} &= \frac{A_{i+1}^{x,n} + A_{i+1}^{x,n+1}}{2};
&
A_{i+1}^{y,n+1/2} &= \frac{A_{i+1}^{y,n} + A_{i+1}^{y,n+1}}{2},
\end{align}
with which, equations \eqref{eq:Ax} and \eqref{eq:Ay} ultimately become:
\begin{align}
\frac{\hh_{i+1}A_{i+1}^{x,n+1/2} - \hh_{i}A_{i}^{x,n+1/2}}{\ff_{i+1/2}}
-
\frac{\hh_{i}A_{i}^{x,n+1/2} - \hh_{i-1}A_{i-1}^{x,n+1/2}}{\ff_{i-1/2}}
&=
- \mu_0 \hh_{i}\ff_{i} j_{i}^{x,n+1/2}
\label{eq:Ax_discrete}
\\
\frac{\hh_{i+1}A_{i+1}^{y,n+1/2} - \hh_{i}A_{i}^{y,n+1/2}}{\ff_{i+1/2}}
-
\frac{\hh_{i}A_{i}^{y,n+1/2} - \hh_{i-1}A_{i-1}^{y,n+1/2}}{\ff_{i-1/2}}
&=
- \mu_0 \hh_{i}\ff_{i} j_{i}^{y,n+1/2}.
\label{eq:Ay_discrete}
\end{align}
This recovers the possibility to compute a stable time-marching solution in the Darwin system with our choice of variable positions in the spatiotemporal grid, and does not affect the charge and energy conservation characteristics of the model \cite{chen14b}.

Lastly, fields at half-integer times are used 
to interpolate the fields at the macroparticles required 
to push them in time, as explained in \ref{sec:macroparticleeqs}. 
With the exception of $E_w^x$ and $E_w^y$, which are already at half-integer times, the remaining fields are interpolated from the fields at integer times via averaging. Thus we define:
\begin{align}
E_{s,i+1/2}^{n+1/2} = \frac{E_{s,i+1/2}^{n} +E_{s,i+1/2}^{n+1} }{2},
\mbox{ etc.}
\end{align}
 
\subsection{Macroparticle equations}\label{sec:macroparticleeqs}

To solve the kinetic Vlasov equation, we introduce the following discretization of $\fdist_\varsigma$ ($\varsigma=i,e$) in terms of markers or macroparticles:
\begin{equation}\label{eq:Dirac}
\fdist_\varsigma\left(\zeta, v_x, v_y, v_z, t\right) = \sum_{p \in \varsigma} w_p \delta\left(\zeta - \zeta_p(t)\right) \delta\left(v^x - v^x_p(t)\right) \delta\left(v^y - v^y_p(t)\right) \delta\left(v^z - v^z_p(t)\right) .
\end{equation}
where $\delta$ denotes the Dirac delta function, and the subindex $p$ runs over all macroparticles corresponding to species $s$. Introducing \eqref{eq:Dirac} into \eqref{eq:vlasov} it is possible to separate the following  evolution equations for each macroparticle:
\begin{align} 
\frac{\dd w_p}{\dd t} &= 0;
&
\frac{\dd \zeta_p}{\dd t} &= \frac{v^z}{f};
&
\frac{\dd v^{x}_{p}}{\dd t} &= a^x;
&
\frac{\dd v^{y}_{p}}{\dd t} &= a^y;
&
\frac{\dd v^{z}_{p}}{\dd t} &= a^z,  
\label{eq:particles}
\end{align}
with $a^x,a^y,a^z$ given by equations \eqref{eq:accelerations}.

The integration of the macroparticle orbits in one $\Delta t$ uses  substepping (i.e., subcycling) for trajectory accuracy when $\Delta t$ is large.
This is done
by breaking $\Delta t$ down into \textit{substeps} of variable duration $\Delta \tau^\nu$, such that $\Delta t = \sum \Delta \tau^\nu$.
This method enables exploiting the difference in scales between the slow field dynamics, represented by $\Delta t$, and the rapid particle dynamics, represented by $\Delta\tau$,
while minimizing memory access, as macroparticle coordinates are stored in fast local registers during subsequent substeps.
The size of each substep, $\Delta\tau^\nu$, is chosen with the timestep estimator of \cite{jime24a}.
On each substep, the macroparticle equations are integrated using a Crank--Nicholson scheme,
\begin{align}
\frac{\zeta^{\nu+1}_p-\zeta^\nu_p}{\Delta \tau^\nu} &= \left(\frac{v^z}{f}\right)^{\nu+1/2}_p,
\label{eq:discretezetaeq}
\\
\frac{v^{x,\nu+1}_p-v^{x,\nu}_p}{\Delta \tau^\nu} &= \frac{q_p}{m_p}\left(E^x_w +v^y B^z_a - v^zB_w^y\right)^{\nu+1/2}_p - \left(\frac{v^xv^z}{\hh \ff}\hh'\right)^{\nu+1/2}_p,
\label{eq:discretevxeq}
\\
\frac{v^{y,\nu+1}_p-v^{y,\nu}_p}{\Delta \tau^\nu} &= \frac{q_p}{m_p}\left(E^y_w -v^x B^z_a+ v^zB_w^x\right)^{\nu+1/2}_p - \left(\frac{v^yv^z}{\hh \ff}\hh'\right)^{\nu+1/2}_p,
\label{eq:discretevyeq}
\\
\frac{v^{z,\nu+1}_p-v^{z,\nu}_p}{\Delta \tau^\nu} &= \frac{q_p}{m_p}\left(E^z_s + v^xB_w^y - v^yB_w^x\right)^{\nu+1/2}_p + \left(\frac{(v^x)^2+(v^y)^2}{\hh \ff}\hh'\right)^{\nu+1/2}_p,
\label{eq:discretevzeq}
\end{align}
where the right hand sides are evaluated in the middle of the substep, i.e., at $\zeta_p^{\nu+1/2}$, $\vec v^{\nu+1/2}$, \rftwo{
\begin{align}
\zeta_{p}^{\nu+1/2} &= \frac{\zeta_{p}^{\nu} +\zeta_{p}^{\nu+1} }{2};
&
\vec v_{p}^{\nu+1/2} &= \frac{\vec v_{p}^{\nu} +\vec v_{p}^{\nu+1} }{2},
\end{align}}
using 
the analytical definition of functions $\hh$, $\ff$.
The fields are interpolated at the macroparticle mid-substep position
from the grid values at time $(n+1/2)$ with the following shape functions, described in appendix \ref{sec:shapefunctions}:
\begin{align}
h_pE_{w,p}^{x,\nu+1/2} &= \sum_i h_iE_{w,i}^{x,n+1/2} S_{i}^2( \zeta_p^{\nu+1/2});
\\
h_pE_{w,p}^{y,\nu+1/2} &= \sum_i h_iE_{w,i}^{y,n+1/2} S_{i}^2( \zeta_p^{\nu+1/2});
\\
f_pE_{s,p}^{z,\nu+1/2} &= \sum_i f_iE_{s,i+1/2}^{z,n+1/2} S_{i+1/2}^1( \zeta_p^{\nu+1/2}).
\label{eq:Ezinterpolated}
\end{align}
These spline orders enable local charge and global energy conservation.
$B_w^x$ and $B_w^y$ are similarly interpolated with $S_{i+1/2}^1$.
Solving the implicit trajectory equations \eqref{eq:discretezetaeq}--\eqref{eq:discretevzeq} requires only a few Picard iterations for each substep when the analytic inversion of the velocity update (implicit-Boris approach \cite{chen14}) is used, which improves numerical conditioning.  

In addition, following the same strategy as in \cite{chen23b,jime24a}, macroparticles are allowed to cross cell boundaries in a given substep $\nu$; in practice, this is done by breaking down each substep into two or more \textit{segments} $s$ (one per cell visited by the macroparticle during the substep), and computing the fields that the macroparticle sees at $(\nu+1/2)$ as a weighted average over the segment midpoints. In other words, this means substituting the spline evaluated at $\nu+1/2$ in equation \ref{eq:Ezinterpolated} as:
\begin{align}
S^m_{i+1/2} (\zeta_p^{\nu+1/2}) \to 
\frac{1}{\Delta \zeta_p^\nu}\sum_\varsigma S^m_{i+1/2}(\zeta_p^{s+1/2})\Delta \zeta_p^s.
\end{align}
A similar substitution needs to be applied in other interpolation and weighting formulas whenever a substep is composed of multiple segments.
Since the Picard iterations are applied on the substep level and not the segment level, there is an important computational saving, demonstrated in \cite{jime24a}, compared to having substeps end at the cell boundaries.

\subsection{Sources weighting}

The computation of the charge current density $\vec\jmath$ at the half-integer time steps
is carried out as follows.
The axial current density $j^z$ is weighted by summing over all substeps of all macroparticles as
\begin{align}
j^{z,n+1/2}_{i+1/2} 
&= \frac{1}{\hh^2_{i+1/2}\Delta t }\sum_{p,\nu} \frac{q_pw_p v_p^{z,\nu+1/2}}{\ff^{\nu+1/2}_p} S^1_{i+1/2} (\zeta_p^{\nu+1/2}) \Delta \tau^\nu_p
\nonumber
\\
&= \frac{1}{\hh^2_{i+1/2}\Delta t }\sum_{p,\nu} q_pw_p S^1_{i+1/2} ( \zeta_p^{\nu+1/2}) \Delta \zeta_p^\nu,
\label{eq:jzweighted}
\end{align}
where, for substeps containing multiple segments, the contribution is substituted by a weighted average over them, analogously as what was explained above.
The charge density $\rho$ is 
defined at the integer positions and integer times as
\begin{align}
\rho^{n}_{i} 
&= \frac{1}{\hh^2_{i}\ff_i}\sum_{p} q_pw_p S^2_{i} (\zeta_p^{n}).
\end{align}
Observe that the computation of the charge density $\rho$ is not required for the advancement of the simulation, as our formulation employs only $\vec \jmath$ (with the exception of the value of $\rho$ on the first and final nodes, which are used to verify the quasineutrality condition there as described in Section \ref{sec:boundaries}).
This form of $j^z$ and $\rho$ discretization ensures local charge conservation per segment \cite{chen11,jime24a}. Indeed, it can be proven that for shape functions up to order 1 (i.e. $S^1$), the midpoint rule used to integrate in equation \eqref{eq:jzweighted} is exact.

On the other hand, the $j^x$ and $j^y$ currents are weighted with $S^2$ shape functions as:
\begin{align}
j^{x,n+1/2}_{i} 
&= \frac{1}{\hh_{i}\ff_i\Delta t }\sum_{p,\nu} \frac{q_pw_p v_p^{x,\nu+1/2}}{h^{\nu+1/2}_p} S^2_{i} (\zeta_p^{\nu+1/2}) \Delta \tau^\nu_p,
\\
j^{y,n+1/2}_{i} 
&= \frac{1}{\hh_{i}\ff_i\Delta t }\sum_{p,\nu} \frac{q_pw_p v_p^{y,\nu+1/2}}{h^{\nu+1/2}_p} S^2_{i} (\zeta_p^{\nu+1/2}) \Delta \tau^\nu_p,
\end{align}
where $h^{\nu+1/2}_p$ and $f^{\nu+1/2}_p$ are here evaluated from the analytical definition of $h$ and $f$ (but could instead be interpolated from the values at the nodes/cell centers if such definition were not available, e.g. in a different problem setting).

At the edges of the domain, care must be put into correctly weighting the moments of the distribution, as the neighboring nodes or cell mid-points only receive (partially) one-sided contributions from the simulated macroparticles. 
This can be taken into account correcting the shape functions of edge and neighboring points by \textit{folding back} into the domain 
any fraction of 
the shape functions of 
appendix \ref{sec:shapefunctions} 
that protrudes beyond the boundary, to maintain the condition that their integral evaluates to 1.
For example,
$S^1_{1/2}(\zeta)$ becomes $\tilde S^1_{1/2}(\zeta)=1$ for $\zeta\in [0,1/2)$, and $\tilde S^1_{1/2}(\zeta)=3/2-\zeta$ for $\zeta\in [1/2,3/2]$.
Indeed, this approach is equivalent to considering ghost macroparticles beyond the domain that are the reflection of those in the domain, and weighting with the unmodified shape functions, as in \cite{erem22}.

\subsection{Boundary and integral conditions}\label{sec:boundaries}

The boundary conditions on the potentials $\phi$ and $\vec A$
described in Section \ref{sec:fields} are applied on the first and last nodes, $i=0$ and $i=N_g$. The value of $\phi_D$ at the downstream boundary is determined as part of the solution as explained below. 

Macroparticles are injected through the upstream boundary of the domain. The sampling of both electrons and ions is carried out from a semi-Maxwellian flux as described in Section \ref{sec:kineticeqs}. 
Injection of each macroparticle is done at a random time instant between $n$ and $n+1$ to avoid artificial particle bunching.
To keep the number of macroparticles per cell ($N_p$) approximately constant and equal to a selected target in the first cell, the weight $w_p$ of the ion and electron macroparticles is varied each time step once the amount of mass to be injected of each species has been determined. As described in \ref{sec:kineticeqs} and further discussed below, the parameter $n_e^*$ of the electron population has to be computed as part of the solution.
Further detail on the injection routines can be found in \cite{jime24a}. 

Macroparticles returning to the upstream boundary are simply removed from the simulation.
On the other hand, macroparticles reaching the downstream boundary are selectively removed or reflected, depending on their energy.
As the potential is assumed to continue to fall monotonically
from its downstream value $\phi_D$
to its infinity value $\phi_\infty<\phi_D$ (also unknown a priori), all ions reaching $D$ are removed.
Electrons, on the other hand, are removed only if their energy allows them to overcome the voltage $\phi_D-\phi_\infty$; otherwise, it is assumed that they would be reflected at some point beyond the end of the domain, and hence, are reflected back at the boundary $D$. \rftwo{This is accomplished by changing the sign of $v^z_p$, and randomizing the phase angle defined by the other two velocity components.}

The values of $n_e^*$, $\phi_D$, and $\phi_\infty$ are computed as part of the solution to satisfy integral conditions on the MN, related with the existence of reflected electrons, and to prevent spurious numerical sheaths at the edges of the domain. 
The balance between the forward and backward electron fluxes must equal the ion current in average for the MN to be globally current free. 
While the ion current everywhere is essentially equal to their injection current, the flux of \textit{forward} electrons is typically much greater than the ion current, and most of them are reflected back by the electrostatic potential $\phi$ at some point within the domain or beyond. 
Only a small fraction of electrons (free electrons) has enough energy to escape to infinity; this fraction is determined by $\phi_\infty$, which must be computed self-consistently.
Moreover, we demand that the density of ions and electrons be equal at the upstream boundary node $i=0$ and  the downstream boundary node $i=N_g$, as the transition from the plasma source to the MN must be quasineutral, and no non-neutralities should develop far downstream; this is affected by $n^*_e$ and $\phi_D$, respectively.

In practice, the temporal update of $n^*_e$, $\phi_D$ and $\phi_\infty$  is accomplished with a simple linear, closed-loop controller that monitors the error in the current at the last cell ($j^z_{N_g-1/2}=0$)
and the quasineutrality at the first  and last nodes ($\rho_0=0$, $\rho_{N_g}=0$), and acts on the value of $n^*_e$, $\phi_D$ and $\phi_\infty$ through a gain matrix $G$.
In the case of $n_e^*$, control is exerted by varying the electron macroparticle weight $w_e$.
In the present work, the matrix is chosen to be diagonal, i.e., 
$n^*_e$ responds only to $\rho_0=0$,
$\phi_D$ to $\rho_{N_g}=0$, and 
$\phi_\infty$ to $j^z_{N_g-1/2}=0$, with the following control laws at every timestep:
\begin{align}
w^{n+1}_e &= w^{n}_e - G_{n_e^*}\frac{n_{eU} - n_{iU}}{n_{eU}};
\\
\phi^{n+1}_D &= \phi^{n}_D -  G_d\frac{T_e^*}{e}\frac{n_{eD} - n_{iD}}{n_{eU}};
\\
\phi^{n+1}_\infty &= \phi^{n}_\infty + G_\infty\frac{\sqrt{m_eT_e^*}}{e^2n_i^*} \bar{\jmath}_z^n.
\end{align}
We empirically find that $G_{n_e^*} = 0.001$, $G_d = 0.005$ and $G_\infty = 1$ enable a speedy convergence to the solution; furthermore, we checked that modifying these values in a wide range \rfone{(i.e. multiplying dividing each gain by a factor of 2---and likely more)} does not change the steady state solution of the plasma or the wave in any noticeable way.

A key aspect of our approach is that it allows obtaining simulation results without artificial sheaths at the upstream or downstream boundaries, which are globally-current free. In addition, this combination of physically-consistent boundary conditions is seen to yield an essentially-quasineutral plasma expansion everywhere.

\subsection{Time marching scheme}

After discretization, a coupled system of nonlinear algebraic equations must be solved for the potentials 
$\mathbf{X_1} = \{\phi^{n+1}_i,\vec{A}^{n+1}_i\}$
and the macroparticles
$\mathbf{X_2} = \{\zeta^{p,n+1}, \vec v^{p,n+1}\}$
at the new time instant $(n+1)$, provided a solution is known at time $n$. 
This is an extremely large number of unknowns.
To make the problem  manageable, Chen et al. \cite{chen11} exploited the fact that the new macroparticle coordinates depend on the potentials, which allows us to express the residual as $\mathbf{R}_1(\mathbf{X}_1, \mathbf{X}_2(\mathbf{X}_1)) = 0$. 
Finding $\mathbf{X}_2(\mathbf{X}_1)$ involves integrating the equations of motion for each macroparticle, given the potentials on the new time instant $\mathbf{X_1}$, and accumulating the moments to form the residual vector $\mathbf{R}$ from the field equations. This process, known as \textit{particle enslavement}, allows a global nonlinear solver such as a Jacobian Free Newton Krylov (JFNK) method, to handle a significantly reduced system of equations without compromising accuracy.

\section{Results}\label{sec:results}

The simulation setup is similar to the one in \cite{jime24a}, albeit with the incident electromagnetic wave and a larger domain size. 
For concreteness, the applied magnetic field on the axis is that of a single current loop of radius $r_L = 1.2\;\mathrm{cm} \gg \ell$:
\begin{equation}
\vec B_a = B_{aT} \frac{r_L^3}{\left(r_L^2+z^2\right)^
{3 / 2}} \vec{1}_z,
\end{equation}
and a magnetic field strength at the throat $T$, found at $z=0$, equal to $B_{aT}=570\; \mathrm{G}$. 
The domain extends from the upstream boundary $0$ at $z = -0.5r_L$ (-0.6 cm) to the downstream boundary $D$ at $z = 17.5r_L$ (21 cm). 
The injection parameters upstream are $n_i^* = 10^{18}$ $\mathrm{m^{-3}}$ and $T_e^*=T_i^*=10$ $\mathrm{eV}$, while $n_e^*$ is determined self-consistently as explained in Section \ref{sec:boundaries}. 

A nonuniform grid with $N_g = 512$ cells is employed, The $\ff$ function is fixed as a fourth-order polynomial 
with the smallest cell size at the entrance $\Delta z = \ff\Delta \zeta =4\lambda_D^*$, and much larger in the rarefied plasma downstream $\Delta z=20\lambda_D^*$ (around $0.3$ local Debye lengths), where $\lambda_D^*=\sqrt{\varepsilon_0T_e^*/n_i^*e^2}=0.0235$ mm is the reference Debye length. 
This choice of $\ff$ dedicates more cells to the upstream region, which features larger potential gradients. Moreover, it approximately matches the evolution of the ion velocity in the solution, which ensures that the number of macroparticles per cell $N_p$ remains approximately uniform throughout the grid as the plasma expands downstream.
Regarding the number of macroparticles per cell $N_p$, in steady state we have about $N_p=3500$--$4000$ everywhere (for each species),
with a total of $3.6$ million macroparticles (both species combined) in the domain. 
The integration time step is fixed to $\Delta t = 7\;(\omega_{pe}^*)^{-1} = 78$ ns,
where $\omega_{pe}^* = \sqrt{n_i^* e^2/(\varepsilon_0 m_e)} = 8.98$ GHz.
Using the substep estimator of \cite{jime24a}, typically 2 substeps per timestep are used. 

A right-hand circularly polarized wave with frequency $\omega/(2\pi)=160\; \mathrm{MHz}$ enters the domain from the upstream boundary. The domain displays an electron-cyclotron resonance (ECR) surface ($\omega_{ce}=\omega$) at position $R$, $z= 2.25$ cm.
To study the effect of the wave on the plasma expansion, we simulate 5 cases with varying input wave power $P_{wU}$: no power (case O), low power: 1.5 $\mathrm{W/cm}^2$ (L), medium power: 7.5 $\mathrm{W/cm}^2$ (M), high power: 37.5 $\mathrm{W/cm}^2$ (H) and extra-high power: 75 $\mathrm{W/cm}^2$ (X). The various simulation cases are presented in table~\ref{tab:steady}.

\begin{table}
\centering
\begin{tabular}{c|c|c|c|c|c|c}
Simulation  & $P_{wU} $ [W] & $\phi_D$ [V] & $\phi_\infty$ [V] & $F_{eD}$ [mN] & $F_{iD}$ [mN]& $\eta_F$ [\%]\\
\hline
O & 0   & -21.30& -29.21& 0.02463& 0.2044& 79.04\\
L & 1.5& -28.03& -31.81& 0.01608& 0.2218& 84.32\\
M & 7.5& -34.20& -38.36& 0.01585& 0.2372& 86.96\\
H & 37.5& -50.01& -60.88& 0.02824& 0.2725& 82.42\\
X & 75& -69.02& -91.93& 0.03903& 0.3061& 79.84\end{tabular}
\caption{Values of the input wave power $P_{wU}$, downstream potential $\phi_D$, potential at infinity $\phi_\infty$,     
ion and electron momentum downstream $F_{iD}$ and $F_{eD}$, and MN efficiency $\eta_F$ (defined in the main text; in particular, $\eta_F$ is defined in equation \eqref{eq:def_etaF}) for the various simulations, normalized for a magnetic tube with a cross-sectional area of $\mathcal A_U=1$ cm$^2$ upstream. 
For all simulation cases, the following values remain essentially unvaried: the reference electron density is $n_e^*= 0.59 n_i^*$; the power of electrons and ions upstream, $P_{eU}$ and $P_{iU}$, is 39 and 19 $\mathrm{W}$, respectively; the momentum flux of electrons and ions upstream, \rftwo{$F_{eU}=0.12$ and $F_{iU}=0.11$ mN}; and the mass flow rate is $\Gamma_i = 5.6\cdot 10^{-10}$ kg/s.
}
\label{tab:steady}
\end{table}

In contrast to other PIC simulations, the physical values of $\varepsilon_0$ and $\mu_0$ 
are kept unchanged.
On the other hand, an ion-to-electron mass ratio $m_i/m_e=100$ is selected
in order to keep results comparable to previous studies \cite{sanc18b,jime24a}.
Simulations were run for a total time $t=31.2$ $\mu$s (for $40,000$ time steps), with steady state reached at about $t=21.8$ $\mu$s. Results shown below are based on the stationary part of the simulations only.
Observe that the problem can be made dimensionless using e.g. $m_e$, $e$, $\lambda_{De}$, $\omega_{pe}$, but dimensional results are presented to facilitate
their interpretation.

\subsection{Steady-state plasma profiles and magnetic nozzle performance}\label{sec:profiles}

\begin{figure}
    \centering
    \includegraphics[width=0.7\textwidth]{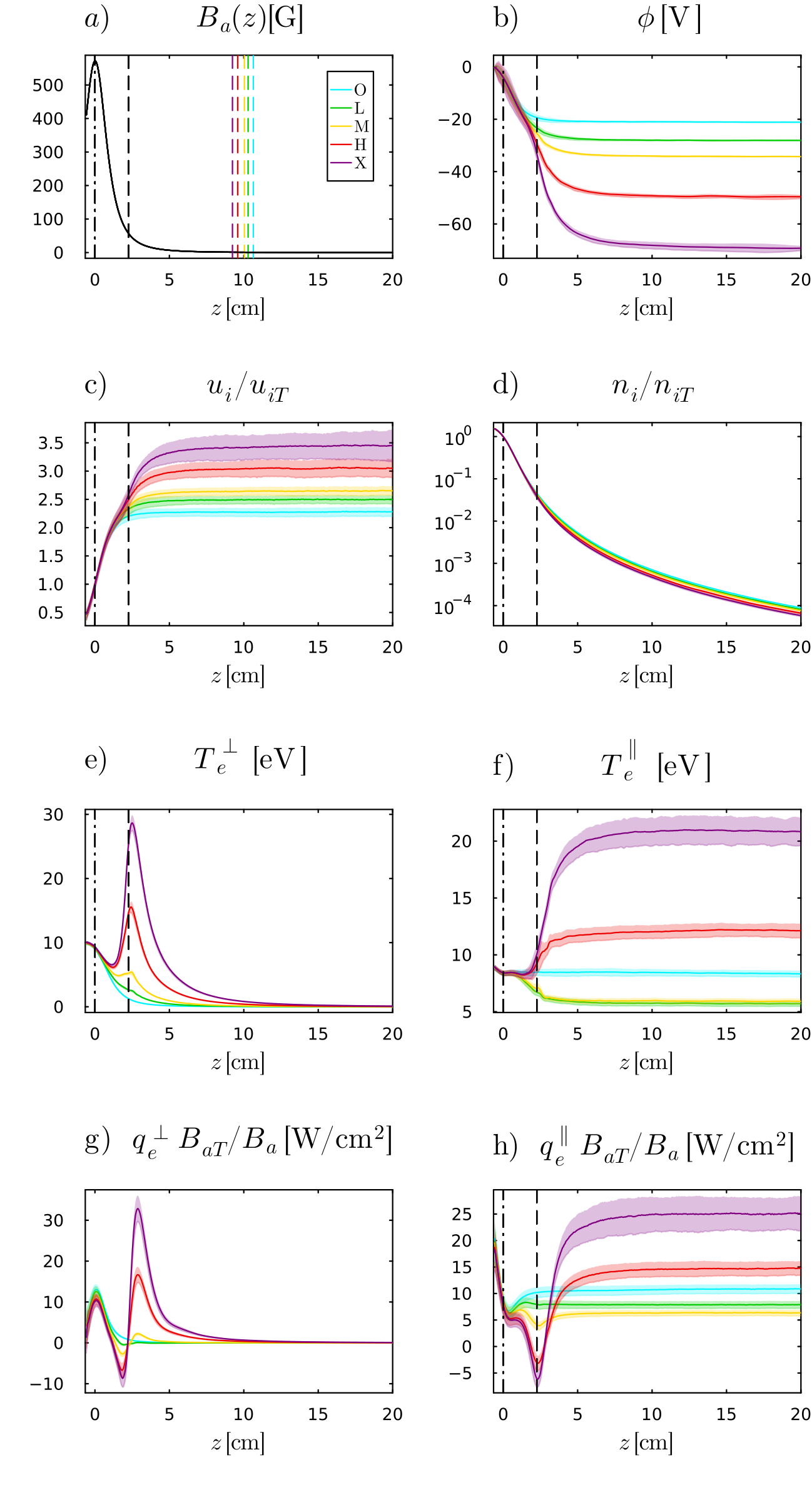} 
    \caption{
    Axial profiles of (a) $B_a$, (b) $\phi$, (c) $u_i$, 
    (d) $n_i$, (e) $T^\perp_{e}$, (f) $T^\parallel_{e}$, (g) $q^\perp_{e}$,  (h) $q^\parallel_{e}$. Colors refer to simulation cases O (cyan), L (green), M (yellow), H (red), X (purple). Solid lines and shaded areas indicate the mean and the standard deviation of the indicated variables over the last $1500$ time steps of the simulation. The black vertical dash-dot and dash lines indicate the location of the throat of the MN ($T$) and the ECR ($R$); the colored vertical dashed lines in (a) correspond to the occurrence of the critical density in each case ($P$).
    }
    \label{fig:steady}
\end{figure}

Figure \ref{fig:steady} shows the steady-state results in the various simulation cases for the electrostatic potential $\phi$, the ion bulk velocity $u_i$, the ion density $n_i$, the electron parallel and perpendicular temperatures $T^\parallel_{e}=p^\parallel_{e}/n_e, T^\perp_{e}=p^\perp_{e}/n_e$, and the parallel flux of parallel and perpendicular electron heat $q^\parallel_{e}, q^\perp_{e}$ (defined in Section \ref{sec:kineticeqs}). For reference, plot \ref{fig:steady}(a) shows the applied magnetic field strength, $B_a$. 
In all simulation cases, off-diagonal terms of the temperature (pressure) tensor are essentially null and are not shown. The bulk electron velocities $u^x_e$ and $u^y_e$ are near-zero everywhere, with the exception of the neighborhood of the resonance $R$, where they are small.

The values of several additional key quantities are given in Table \ref{tab:steady} and its caption.
Some of these are related with the ion and electron mass, momentum, and energy fluxes, $\Gamma_\varsigma$, $F_\varsigma$ and $P_\varsigma$ ($\varsigma=i,e$) in a tube of cross section $\mathcal A(\zeta) = \mathcal A_U h^2(\zeta)$, which are defined as:
\begin{align}
\Gamma_\varsigma(\zeta) &= \underbrace{m_\varsigma n_\varsigma u^z_\varsigma \mathcal A}_{\gamma_{\varsigma1}},
\label{eq:massflowdef}
\\
F_\varsigma(\zeta) &= \underbrace{m_\varsigma n_\varsigma (u^z_\varsigma)^2 \mathcal A}_{\varphi_{\varsigma 1}} + \underbrace{p^\parallel_\varsigma \mathcal A}_{\varphi_{\varsigma 2}},
\label{eq:momentumflowdef}
\\
P_\varsigma(\zeta) &=
\underbrace{\frac{1}{2}m_\varsigma n_\varsigma (u_\varsigma)^2u^z_\varsigma \mathcal A}_{\pi_{\varsigma 1}}
+ 
\underbrace{p^\perp_\varsigma u^z_\varsigma \mathcal A}_{\pi_{\varsigma 2}}
+
\underbrace{\frac{3}{2}p^\parallel_\varsigma u^z_\varsigma \mathcal A}_{\pi_{\varsigma 3}} 
+
\underbrace{q^\perp_\varsigma \mathcal A}_{\pi_{\varsigma 4}} 
+
\underbrace{q^\parallel_\varsigma \mathcal A}_{\pi_{\varsigma 5}},
\label{eq:energyflowdef}
\end{align}
where the names to each contribution will be used in Section \ref{sec:balances}.
For definiteness, we have taken $\mathcal{A}_U = 1$ cm$^2$ in the table. The thrust efficiency of the MN at station $D$ is defined as 
\begin{align}
\eta_F = \frac{(F_{iD}+F_{eD})^2}{2\Gamma_i(P_{iU}+P_{eU}+P_{wU})},
\label{eq:def_etaF}
\end{align}
where we have dropped $\Gamma_e\ll\Gamma_i$ in the denominator.

The potential $\phi$ in Figure \ref{fig:steady}(b) decreases axially, quickly at first, driving the acceleration of ions at the expense of the electron energy. 
Indeed, this potential fall forms self-consistently to confine the majority of the electrons, with a thermal velocity much larger than that of ions, so that the plasma jet is effectively current-free, and it plays the essential role of converting electron thermal energy to directed ion energy \cite{ahed10f}.
Results evidence that the drop in $\phi$ becomes larger as the wave power $P_{wU}$ is increased, consistent with a higher electron energy being available in the domain due to electromagnetic power deposition. 
The potential fall at infinity in table \ref{tab:steady}, $\phi_\infty$, also scales with the available electron energy. 
A second quick decrease in $\phi$ takes place at the resonance, especially in the simulations with higher $P_{wU}$. 

The response of $u_i$ and $n_i$ is depicted in Figure \ref{fig:steady}(c--d).
As in any plasma expansion to vacuum, bulk ion velocity increases as the potential falls due to ion energy conservation. 
The overall larger potential fall in the simulations with $P_{wU}$, in turn, drives a larger $u_i$. 
The density $n_i$ decreases monotonically downstream. This is mostly the effect of the expanding cross section of the magnetic tube. The different $u_i$ profiles across simulations explain the small differences in the $n_i$ profiles, as per the conservation of the particle flux in equation \eqref{eq:massflowdef}.  
All simulations are seen to be essentially quasineutral everywhere in the domain, with the typical value of $(n_i-n_e)/n_i$ being around~$10^{-3}$.

In the electrostatic scenario (simulation O), the magnetic moment of each particle, $\mu = m(v^{\perp})^2/(2B_a)$ (with $(v^{\perp})^2=(v^x)^2+(v^y)^2$), is conserved, so that the electron perpendicular energy  responds to changes in the magnetic field strength  only (i.e., the tube cross sectional area). Indeed, in the divergent side of the MN, $T^\perp_{e}$ decreases almost linearly with $B_a$. The magnetic mirror force acts behind this reduction, converting the perpendicular electron energy to axial energy. 
In the electromagnetic simulation cases, this behavior is modified at and around the resonance region, where a net increase in the  perpendicular electron temperature takes place. Heating is caused by the power transfer from the wave to the electrons, which is further explored in Section \ref{sec:fields}. However, downstream of the ECR, where the wave fields are negligible, the trend $T^\perp_{e} \sim B_a$ resumes. 
The existence of a local peak of $T^\perp_{e}$ is also reported in  previous experimental and numerical studies of  the plume of HPTs \cite{jime23a,vinci23a}, where it was seen that local heating takes place near the ECR surface that exists within the~MN.

The larger $T^\perp_{e}$ in the divergent side in the electromagnetic cases effectively causes a larger rightward magnetic mirror push on the electrons, which is countered by the larger electrostatic potential fall attested by figure \ref{fig:steady}(b) and the value of $\phi_D$, $\phi_\infty$ in table \ref{tab:steady}. 
The more energetic electrons demand a lower $\phi_\infty$ to guarantee the MN remains current-free, by ensuring the right amount of electrons is reflected back toward the source.

Previous studies \cite{sanc18b,jime24a} (and our electrostatic case O) indicate that in the absence of any wave heating, the MN tends to keep the electron parallel temperature $T^\parallel_{e}$ fairly constant throughout the expansion. 
In contrast to $T^\perp_{e}$, temperature $T^\parallel_{e}$ does not go to zero downstream, as the escaping free electrons possess and maintain some velocity spread.
However, when the wave heats the electrons in the perpendicular direction, a different structure can emerge. This is the direct consequence of the modification of the free electron distribution by the wave: the extra perpendicular energy gained at the resonance is converted to parallel energy by the magnetic mirror effect. This raises the cut off energy $-\phi_\infty$ that delimits the free electron region in phase space, but also changes their phase space density, as some low energy electrons are pumped into this region. 
The net effect on $T^\parallel_e$ differs for low and high values of the input wave power, $P_{wU}$:
At low wave powers, the downstream value of $T^\parallel_{e}$ actually decreases, up to a certain threshold in $P_{wU}$. 
Beyond that, the trend reverses, and the downstream $T^\parallel_{e}$ grows with further increase of $P_{wU}$. The underlying reason for this behavior stems 
from how `concentrated' the electron velocity distribution function becomes in the $v^\perp$ direction at the resonance, as ultimately 
the perpendicular-to-parallel energy conversion enacted by the MN will bring this spread into the $v^z$ direction, and thus affect $T^\parallel_e$. At low wave powers, the spread in the distribution in $v^\perp$  at the resonance continues to be small, and the distribution in $v^z$ far downstream is similar to the one of the electrostatic case, although slightly skewed, with the net result that $T^\parallel_e$ decreases somewhat. At higher wave powers, however, the wave has enough authority to concentrate the distribution in $v^\perp$ away from $v^\perp=0$. When this is converted into the axial direction by the mirror force, the result is a distribution in $v^z$ with a greater spread, and thus higher $T^\parallel_e$.
This non-trivial evolution of the electron velocity distribution function $\fdist_e$ is further inspected in Section \ref{sec:evdf}, and has a direct consequence on the performance of the device as discussed further below.

The heat fluxes $q^\perp_{e}$ and $q^\parallel_{e}$ measure the asymmetry in the tails of the electron distribution function $\fdist_e$, namely, the part of it associated with free electrons. 
Figure \ref{fig:steady}(g) shows that $q^\perp_e$ is essentially zero at the ECR surface $R$, and is negative to the left and positive to the right of it. This is consistent with a heat flux emanating from the region where wave power is absorbed and away from it in both directions. The leftward heat flux is due to returning, left-traveling electrons that pass again through the resonance. The effect is more marked the higher $P_{wU}$.

Similarly to $T^\parallel_e$, the behavior of the heat flux of parallel energy, $q^\parallel_e$, shown in \ref{fig:steady}(h), is nonmonotonic with $P_{wU}$.
Under no or small wave power, $q^\parallel_e$ remains small, and decreases with $P_{wU}$; as wave power increases beyond certain value, though, $q^\parallel_e$ increases downstream with $P_{wU}$. The heat flux $q^\parallel_e$ becomes negative at $R$ at higher wave powers, coincident with this change of trend.
The contribution of $q^\parallel_e$ to the electron power flux remains frozen downstream in the plasma jet, and is unavailable for further ion acceleration. Therefore, it counts as a loss term.
Our results evidence that high $P_{wU}$ leads to increased electron heat losses.

\begin{figure} 
    \centering
    \includegraphics[width=0.55\linewidth]{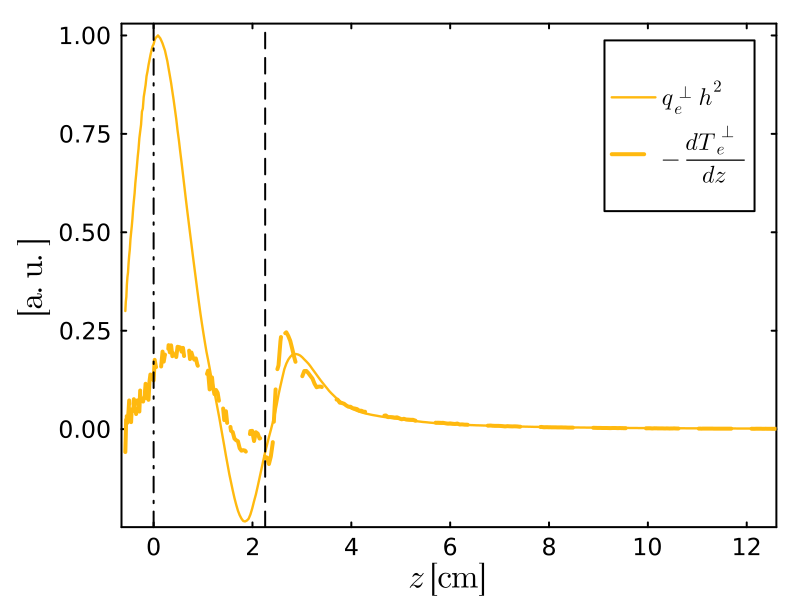}
    \caption{
    Arbitrary-unit comparison of $q^\perp_e$ (solid line) and $-\dd T^\perp_e/\dd z$ (dashed line) for simulation case M. The black vertical dash-dot and dash lines indicate the location of the throat of the MN ($T$) and the ECR ($R$).
    } 
    \label{fig:heatfluxdiscussion}
\end{figure}

Interestingly, the shape of $q^\perp_e$ in the MN is somewhat reminiscent of that of $-\dd T^\perp_e / \dd z$ (figure \ref{fig:heatfluxdiscussion}); \rfone{qualitatively,} this suggests that a Fourier-like heat flux law \rfone{might} be able to approximate its evolution empirically, even in this collisionless setting where such mathematical expression need not apply. The same cannot be stated of $q^\parallel_e$ and $-\dd T^\parallel_e / \dd z$, however, which denotes that no simple fluid-like closure for $q^\parallel_e$ is apparent; a deeper discussion of fluid closures in the electrostatic case can be found in \cite{ahed20a}.

An interesting observation is that none of the variables shown in figure \ref{fig:steady} is much affected by the wave power in the upstream part of the domain, where the electron distribution $\fdist_e$ is dominated by the unadulterated injected electrons and the reflected electrons that find their turning point prior to station $R$. As soon as the injected electrons traverse the resonance, however, their population is drastically modified, especially at higher wave powers.
The effect of the power absorption on the plasma properties to the left of the resonance is small and limited in extent, as returning electrons passing through the resonance for a second time typically acquire a large magnetic moment, and thus are prevented from crossing to the convergent side 
or even approaching the neighborhood of the MN throat 
by the magnetic mirror effect.

Finally, we note the remarkable smallness of the standard deviation achieved by the numerical scheme in $\phi$, $u^z_i$, $n_i$, $T^\perp_e$, and $q^\perp_e$. This is in good part due to the large count of particles per cell $N_p$.
The standard deviation of $T^\parallel_e$ and $q^\parallel_e$ (and also that of $u^z_e$, not shown), albeit also acceptable when compared to other PIC codes, shows a larger spread, especially at higher wave powers. This is symptomatic of these parallel variables depending more strongly on the free electron population, whose distribution is less resolved than that of the trapped electrons. In addition, as in all full-$\fdist$ particle codes, the statistical error increases with the order of the fluid moment.

Coming back to the performance figures in table \ref{tab:steady}, 
it is interesting to note that, albeit the contribution of electron pressure to thrust at the end of the domain, $F_{eD}$, is always negligible compared to the ion thrust, $F_{iD}$, it follows the trend of $T^\perp_e$ observed in figure \ref{fig:steady}(e): up to mid wave powers, it decreases with respect to the electrostatic case, and for higher powers, it increases again past that value. Indeed, $F_{eD}$ is essentially the electron pressure force at the end of the domain.
In contrast, the ion thrust $F_{iD}$ increases monotonically with the extra power.

The thrust efficiency associated with the MN,  $\eta_F$ \rfone{at $D$}, as defined in equation \eqref{eq:def_etaF}, is the key performance figure of the MN under the effect of the electromagnetic wave.  
\rfone{It measures the useful propulsive power in the plasma downstream, $(F_{iD}+F_{eD})^2/2\Gamma_i$, against and all the power entering the MN (ions, electrons, and wave). Observe that by measuring this quantity at D, we do not account for any further changes in the plasma fluxes between there and infinity, although these have asymptoted already inside the domain.}
Incidentally, we observe that the input ion and electron power and momentum at $U$, i.e. $P_{iU}$ and $P_{eU}$, are essentially unchanged across the simulations, consistently with the particle distributions not being much affected by the wave to the left of the resonance. 
\rfone{
Accepting that ion pressure and electron inertia are negligible downstream, and that $F_{eD} \ll F_{iD}$, 
then $(F_{iD}+F_{eD})^2/2\Gamma_i \simeq \pi_{i1}+\pi_{e3}$, the energy flux terms that represent useful propulsive power. Then, the main loss term is indeed $\pi_{e5}$, the heat flux of parallel energy of electrons.}
Our results 
show that when the wave power seeping into the MN from the source is small,  a beneficial albeit limited increase in the overall efficiency can occur.
This holds until the decreasing trend of $T^\parallel_e$ and $q^\parallel_e$ is reverted: at higher wave powers, the extra losses associated with higher $T^\parallel_e$ and $q^\parallel_e$ \rfone{(and thus $\pi_{e5}$)} begin to pull down $\eta_F$, in spite of the monotonically increasing ion thrust $F_{iD}$.

\begin{figure}
    \centering
     \includegraphics[width=0.7\linewidth]{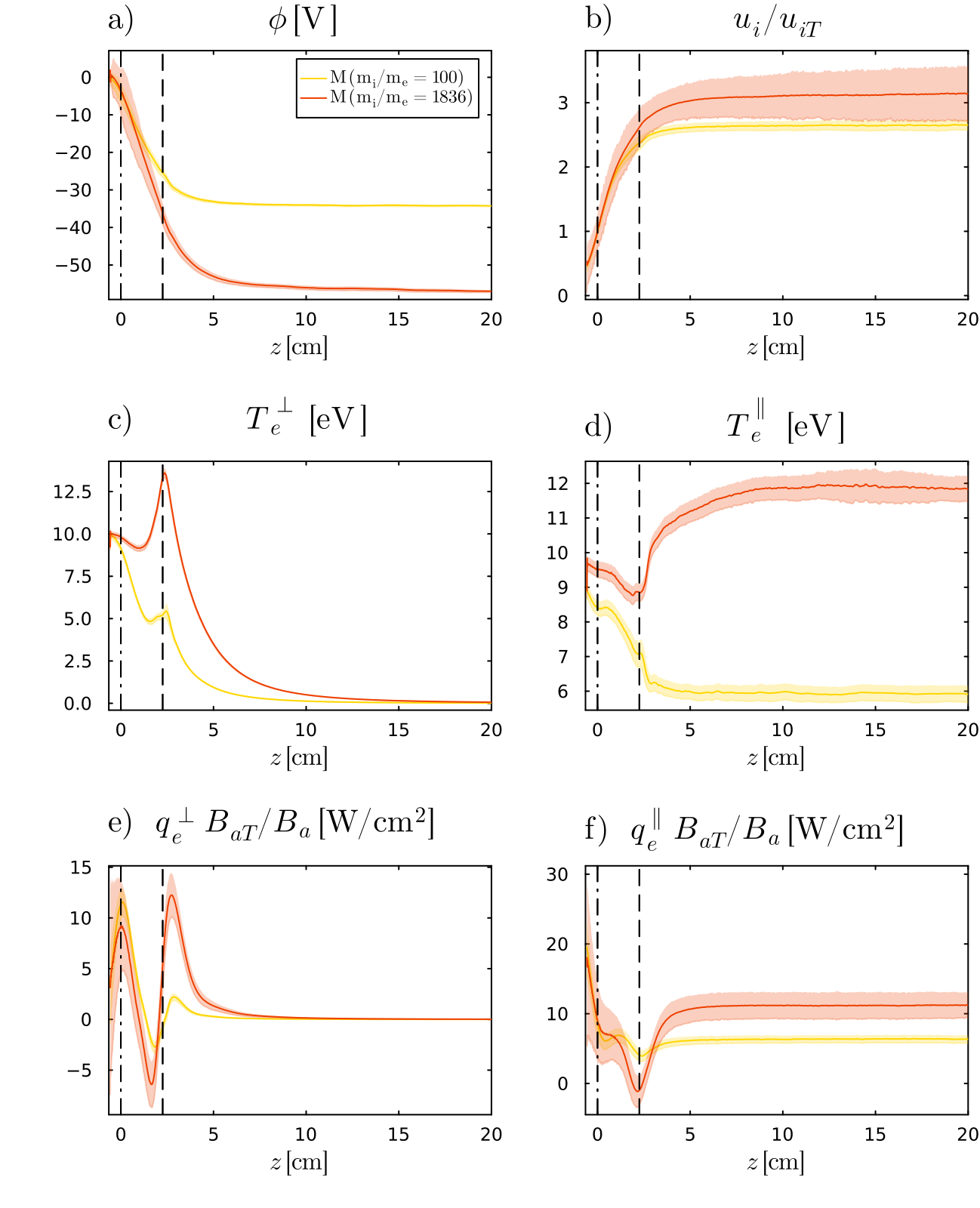}
    \caption{\rfone{
    Axial profiles of (a) $\phi$, (b) $u_i$, (c) $T^\perp_{e}$, 
    (d) $T^\parallel_{e}$, (e) $q^\perp_{e}$, (f) $q^\parallel_{e}$ for the additional simulation with a mass ratio of $m_i/m_e= 1836$ (orange).
    The reference simulation M (with $m_1/m_e=100$) is shown in yellow for comparison.
    Solid lines and shaded areas indicate the mean and the standard deviation, computed over the last $1500$ time steps of the simulation. The black vertical dash-dot and dash lines indicate the location of the throat of the MN ($T$) and the ECR ($R$).
    }
    }
    \label{fig:massratio}
\end{figure}

\rfone{We close this section by briefly noting the effect of the ion-to-electron mass ratio, $m_i/m_e$ on the simulation results. An additional simulation was run for hydrogen ions ($m_i/m_e=1836$ rather than $100$) with $P_{wU}=7.5$ W (same wave power as simulation M). Results are shown in figure \ref{fig:massratio}.
As known from previous kinetic works without wave power (see e.g. \cite{jime24a} and, in particular, figure 10 therein), a larger $m_i/m_e$ results in a larger potential fall. Associated with this, the fraction of trapped electrons increases, which drives the response of this species. For example, electron temperature rises, as electrons become isothermal and Boltzmann-like in the limit of a perfectly-confined electron population.
These trends persist under the influence of wave power. Potential drops to $\phi_D=-57$ V rather than $\phi_D=-34$ V. 
Interestingly, the electron temperature and heat flux response is akin to those of the `high-$P_w$' cases, which suggests that the threshold $P_{wU}$ value that separates the low- from the high-$P_w$ regimes depends on $m_i/m_e$, becoming lower as this parameter increases.
Computationally, the only impact of increasing $m_i/m_e$ is the longer physical time needed to reach steady state, and therefore the longer wall time, for the same $\Delta t$.
}

\subsection{Wave fields and power deposition}\label{sec:wave_fields}

\begin{figure}
    \centering
    \includegraphics[width=\textwidth]{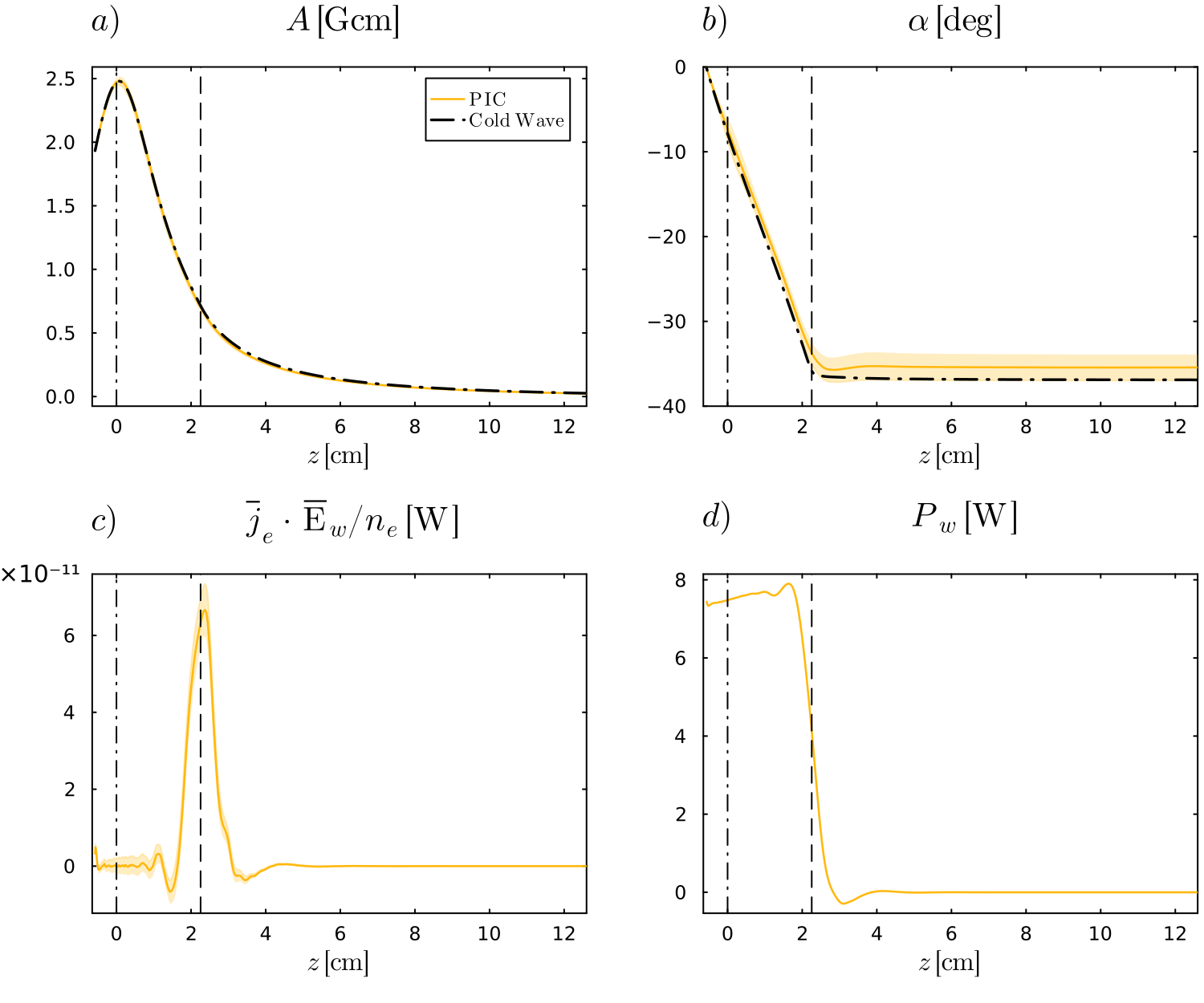}     \caption{
    Electromagnetic wave results for simulation case M: (a) Magnitude $A=\sqrt{A_x^2+A_y^2}$ and (b) phase angle $\alpha = \arctan(A_y/A_x)$, refered to section $U$ upstream, compared with cold wave approximation (fitted cold plasma model); (c) Mean power absorption per electron; (d) Poynting flux. Solid lines
    and shaded areas indicate the mean and the standard deviation, computed over the last
    1500 time steps. {In the case of the phase angle, averaging is done by first subtracting the phase of the first node.} The black vertical dash-dot and dash lines indicate the location of the throat of the MN ($T$) and the ECR ($R$).}
    \label{fig:fields}
\end{figure}

The magnitude and phase of the wave potential $\vec A$, the wave power absorbed per electron, and the wave power balance for simulation M are shown in Figure \ref{fig:fields}.  
Geometric scaling and standard WKB theory 
show that, in the absence of any power exchange with the plasma and away from the resonance, the magnitude $A$ of a traveling wave evolves to first approximation as $A^2\propto 1/(kh^2)$.
While $k$ varies mildly upstream of the resonance, $h$ does change significantly, reaching a minimum at the throat $T$, which explains the maximum of $A$ near this position.  
Further downstream and around the resonance $R$, $A$ decreases faster than this scaling, corresponding to wave power being transferred into the plasma. Beyond the resonance, the field amplitude does not vanish immediately, but gradually decays to zero as a combined effect of the evanescence in this region and the expansion of the MN (i.e., as $h$ grows).
The wave propagates into the MN up to the resonance $R$ with a nearly constant phase rate as shown in figure \ref{fig:fields}(b), completing about 1/10th of a wavelength in this first part of the domain. Beyond that point, it becomes evanescent and the phase angle remains essentially flat. This continues until the end of the domain.

The absorption of wave power per electron, shown in figure \ref{fig:fields}(c), evidences 
the concentrated absorption that takes place at and near the resonance, which explains the behavior of the plasma properties described in Section \ref{sec:profiles}.
The non-zero width of the absorption region is related with the transit-time resonance width, which is $[2\pi m v^\parallel_e/(e\dd B_a/\dd z)]^{1/2}$, about 1 cm for thermal electrons with $v_e^\parallel = \sqrt{T_e/m_e}$.
Doppler broadening also contributes to increasing the width of the absorption peak; however, a quick estimate shows that the Doppler displacement of the resonance for the typical electron velocities is smaller than the peak half-width: around $1.5$ mm for thermal electrons.

Interestingly, the power exchange goes both ways: around this absorption region there are two smaller ones where some net power flows from the plasma into the wave.
A possible explanation of this is the process by which the gyrophase of electrons approaching the resonance, which is initially uniformly distributed, gradually synchronizes with the wave field phase; in this process, some of the electrons gain energy from the fields while others lose it to them. Once the synchronization (or alignment) is essentially complete, all electrons can gain energy efficiently from the resonating fields. The evolution of the electron gyrophase distribution is further discussed in Section \ref{sec:evdf} where it is shown that this alignment does indeed take place.

A complementary view into power absorption is provided by
the Poynting flux $\mathcal S$.
In analogy to expressions \eqref{eq:massflowdef}--\eqref{eq:energyflowdef} we define the flux of electromagnetic energy
in the magnetic tube as
\begin{align}
P_w(\zeta) =
\mathcal{S}(\zeta) \mathcal A(\zeta)  
= 
\frac{1}{\mu_0}\left(E_w^xB_w^y - E_w^yB_w^x\right)\mathcal A 
\label{eq:poyntingdef}
\end{align}
where again $\mathcal A(\zeta) =\mathcal A_U h^2$, and we fix $\mathcal A_U = 1$ cm$^2$ for concreteness. To perform this computation, the wave magnetic field $\vec B_w$, which is computed at half-integer positions, is interpolated at the nodes. Similarly, the electric field $\vec E_w$ at integer timesteps is used.
This is shown in Figure \ref{fig:fields}(d), which evidences that the wave energy is consumed around the resonance $R$ and that the wave power in the evanescent region is essentially zero.

Finally, it is worthwhile to compare these results with the  cold plasma dielectric tensor theory \cite{STIX92}, which has been used in the past to model HPTs \cite{jime23a} and ECRTs \cite{svil21a,svil23}. Within that approximation, the amplitude of a right-hand polarized wave
along a slowly expanding magnetic tube, with varying plasma density, is given by 
\begin{align}
\partial_\zeta\left[\frac{1}{f}\partial_\zeta(hA)\right] + k_R^2 \hh \ff A=0
\label{eq:coldplasmawaveeq}
\end{align}
where $A(z)\exp(-i\omega t) = A^x(z,t) - i A^y(z,t)$ and
$k_R$ is essentially the same as in equation \eqref{eq:kR}, although now we include \rfone{an additional free parameter, the} collisional term $\nu$, to enable power absorption by the plasma:
\begin{align}
k_R^
2(\zeta) = \frac{\omega^2}{c^2}\left(1 - \frac{\omega_{pe}^2(\zeta)}{\omega[\omega-\omega_{ce}(\zeta) + i\nu]}\right).
\label{eq:kRcoll}
\end{align} 
Observe that the first term in the parenthesis is kept, consistent with the full-Maxwell model from which the cold plasma model originates (this is dropped in the Darwin model).
Entering the upstream plasma properties from our simulation and $\nu=0$, the cold plasma wavelength in the upstream region is around $2\pi/k_R\simeq 25$ cm. This matches well with the roughly 1/10th of wavelength seen in the kinetic model from station $U$ to station $R$.

\rfone{Once a value (or a map) of $\nu$ is provided}, the differential equation \eqref{eq:coldplasmawaveeq} is readily solved to meet the boundary conditions associated with our problem: a chosen wave power entering the domain from the left, and no wave power input from the right. 
In order to obtain a good match with the kinetic-electromagnetic result, it was necessary to carefully tune the effective collisionality parameter.
Indeed, keeping $\nu\to 0$ or excessively large gives results that are widely off from the kinetic simulation. After tuning, a  value
$\nu=8.25 \cdot 10^{-4}\omega_{pe}^* = 7.40 \; \mathrm{MHz}$ was found to deliver a good match in this particular simulation case. 
\rfone{The best fit of this model is superimposed in figure \ref{fig:fields} for comparison with the kinetic simulation.}
However, our tests show that this value needs to be adjusted for each different scenario, highlighting a first major limitation of the cold plasma model as a surrogate model.

It should be noted that the plasma cutoff near position $P$ does not play any role \rfone{in the problem under analysis} in our
Darwin electromagnetic approximation: the cutoff simply does not take place  due to the $1$ unit shift in the squared dimensionless refractive index. 
In fact, $P$ does not play any significant role in the full-Maxwell cold plasma model either, where it delimits the actual transition from evanescence into a new propagation region, albeit with a very small $k$. 
Indeed, in both models the phase angle in figure \ref{fig:fields}(b) remains essentially constant from the resonance $R$ to the end of the domain $D$; any variation of the phase after the cutoff occurs in a much longer length scale than the size of the domain, 
comparable to the large local wavelength in that region.
This indicates that, although the Darwin approximation loses accuracy where the refractive index is not $\gg 1$, the error committed in the downstream region is harmless in the present application.
 
\subsection{Conservation laws and dominant balances}\label{sec:balances}

In addition to inspecting individual moments of the particle distributions $\fdist_i$ and $\fdist_e$ above, further insight can be gained by inspecting the mass, momentum, and energy conservation laws of each species term by term, to determine what the dominant terms are in each balance equation.
In our magnetic tube, these can be written in compact form using the mass, momentum, and energy fluxes defined in equations \eqref{eq:massflowdef}--\eqref{eq:energyflowdef} as:
\begin{align}
\gamma_\varsigma(\zeta) - \gamma_{\varsigma U} &= 0,
\\
\varphi_\varsigma(\zeta) - \varphi_{\varsigma U} &= \mathcal{A}_U\Bigg[
\underbrace{\int_0^\zeta q_\varsigma n_\varsigma E^z_s \hh^2  \ff\dd \zeta}_{M_{\varsigma 1}}
+\underbrace{\int_0^\zeta (j^x_\varsigma B_{w}^y - j^y_\varsigma B_{w}^x) \hh^2 \ff\dd \zeta}_{M_{\varsigma 2}}
\nonumber
\\
&+ \underbrace{\int_0^\zeta 
\left[p^{\perp}_\varsigma +\frac{1}{2}m_\varsigma n_\varsigma(u^\perp_\varsigma)^2\right]  2\hh \hh' \dd \zeta}_{M_{\varsigma 3}}\Bigg],
\label{eq:mombalance}
\\
\pi_\varsigma(\zeta)-\pi_{\varsigma U} &= \mathcal A_U \Bigg[
\underbrace{\int_0^\zeta j^z_\varsigma E^z_s \hh^2 \ff \dd \zeta}_{H_{\varsigma 1}}
+\underbrace{\int_0^\zeta (j^x_\varsigma E^x_w + j^y_\varsigma E^y_w) \hh^2 \ff \dd \zeta}_{H_{\varsigma 2}}\Bigg].
\label{eq:energybalance}
\end{align}
Again, we pick $\mathcal{A}_U=1$ cm$^2$ for concreteness.
Incidentally, the first integral in equation \eqref{eq:energybalance} can be integrated by parts to yield $H_{\varsigma 1} = -\mathcal{A}_U[j^z_\varsigma(\zeta)\phi(\zeta)\hh^2(\zeta) - j^z_\varsigma(0)\phi(0)$].

\begin{figure}
\centering
\includegraphics[width=1\textwidth]{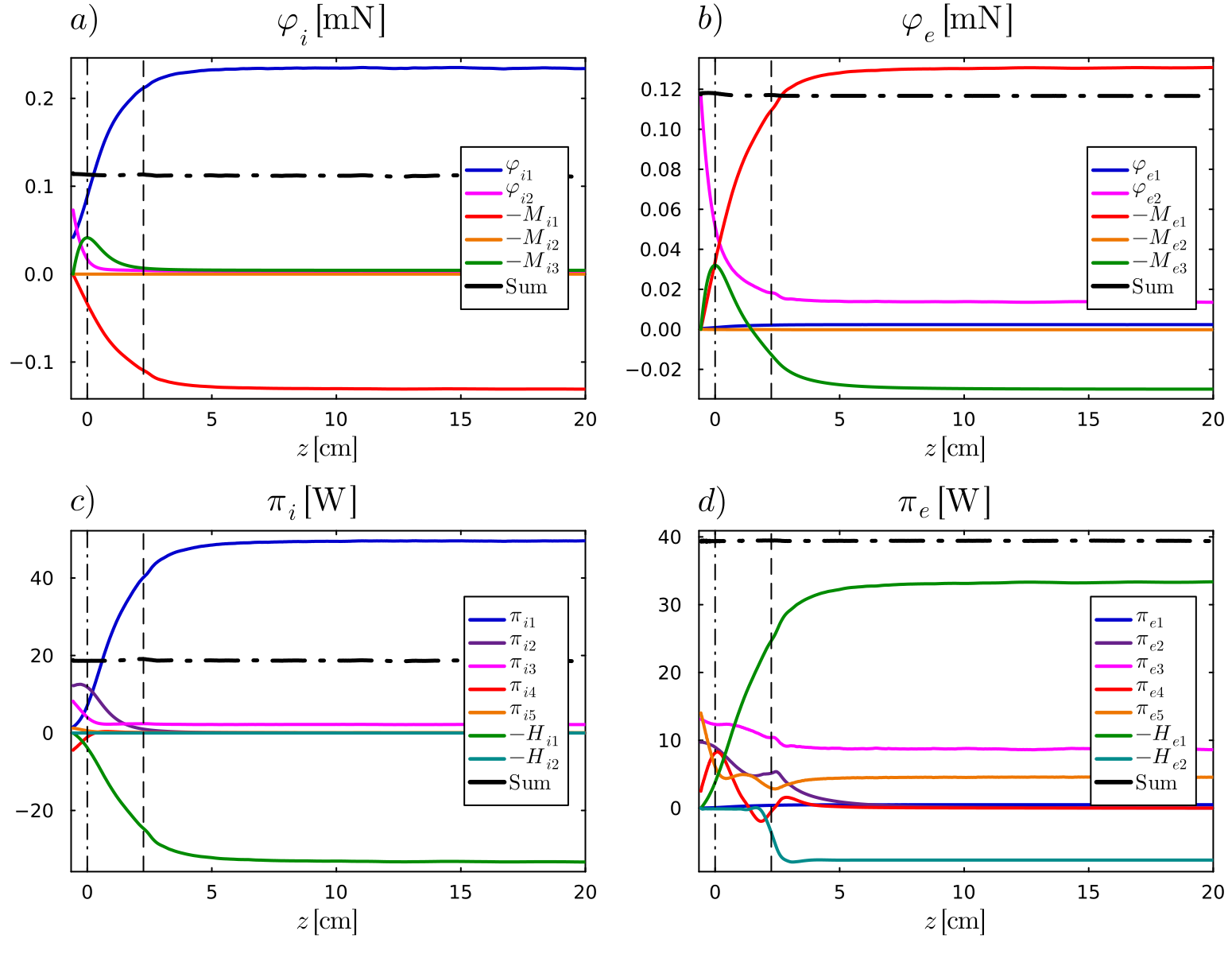}
\caption{ Momentum and energy equation balances for electrons (b and d) and ions (a and c), for simulation case M.
\rftwo{The definition of the $\varphi_\varsigma$, $\pi_\varsigma$ quantities is given in equations
\eqref{eq:momentumflowdef} and \eqref{eq:energyflowdef}. The definition of the $M_\varsigma$, $H_\varsigma$ quantities is given in equations \eqref{eq:mombalance} and \eqref{eq:energybalance}.}
\rfone{The lines correspond to these quantities in a reference magnetic tube with initial area upstream of $\mathcal A_U = 1$ cm$^2$.}
The black vertical dash-dot and dash lines indicate the location of the throat of the MN ($T$) and the ECR ($R$). Quantities have been averaged over the last 1500 time steps. }
\label{fig:balances}
\end{figure}

Figure \ref{fig:balances} displays the evolution along the MN of the $\varphi_\varsigma$, $\pi_\varsigma$ fluxes, and the $M_\varsigma$, $H_\varsigma$ volume integrals, for ions and electrons ($\varsigma=i,e$). 
The global mass balance is not shown nor discussed, as it is satisfied exactly to machine precision as in any mass-conserving PIC code.
Plot (a)  shows that 
directed ion momentum $\varphi_{i1}$ increases, while the parallel pressure term $\varphi_{i2}$ decreases to a residual value as the plasma expands. 
The net growth of total ion momentum is enabled by two mechanisms: the main one is electrostatic acceleration $M_{i1}$, but since in our simulations $T_i^* = T_e^*$, ion pressure is non-negligible, and there is also a contribution from the conversion of the perpendicular ion pressure by the magnetic mirror force in \rftwo{$M_{i3}$}. This second term is actually a deceleration force until the MN throat.

In the electron momentum  in figure \ref{fig:balances}(b), the inertial term $\varphi_{e1}$ is completely negligible. Just like with the ions, the parallel pressure term  $\varphi_{e2}$ decreases to a residual value.
The electrostatic force integral $M_{e1}$
decelerates electrons and confines all of them except for the higher energy, free electrons.
It balances out the second integral term $M_{e2}$ that develops $p^\perp_e$ into axial momentum thanks to the magnetic mirror force; like with $M_{i2}$, this term is a negative contribution before the MN throat.
Since $M_{e1}=-M_{i1}$, a key aspect of the electrostatic field $E^z_s$ is that it helps transfer the  momentum delivered by $M_{e2}$ into the ions. Indeed, this is a main mechanism in the operation of electron-driven MNs: the mirror term spends the perpendicular electron pressure to generate axial electron momentum, and simultaneously, the electrostatic force converts it into axial ion momentum.

The ion energy balance in figure \ref{fig:balances}(c) confirms that indeed the dominant power balance is the conversion of electrostatic energy into directed kinetic energy ($\pi_{i1}$) via the term $H_{i1}$. The remainder flux and integral terms are small and only play a relevant role in the convergent part of the MN. The wave does not interact energetically with the ions to any significant extent.

Finally, the most interesting balance is that of the electron energy equation in figure \ref{fig:balances}(d).
The electrostatic integral $H_{e1}$ is equal and opposite to $H_{i1}$, making it again manifest that the energy of the electrons is used to fuel the ion acceleration, mediated by the electrostatic field.
The wave field integral $H_{e2}$ shows the power deposition into the electrons that takes place around the resonance; the power absorption coincides with the power lost by the wave, shown in figure \ref{fig:fields}. This extra power input to the electrons is spent mainly into raising the perpendicular pressure energy flux term, $\pi_{e2}$. 
All this extra power in $\pi_{e2}$ is eventually converted to parallel energy forms thanks to the magnetic mirror effect.
It is also evident that the energy flux associated with the parallel pressure, $\pi_{e3}$, is kept high thanks to the conversion of some random energy from the perpendicular direction to the parallel direction by the mirror effect.
Finally, heat fluxes $\pi_{e4}$ and $\pi_{e5}$ are seen to be non-negligible, especially for the parallel heat term $\pi_{e5}$. This is true even in the absence of a wave in case O (not shown), and it reflects the asymmetry existing in the electron distribution function $\fdist_e$, where there is no counterpart to the free electron population in the $v^z<0$ side of phase space, as further explained in Section \ref{sec:evdf}.

In each plot, the balance of the participating terms is indicative of the accuracy of the steady state solution. The largest errors in each are $2.99\%$ and $2.45\%$ for the ion and electron momentum balances, and $1.23\%$ and $0.28\%$ for the ion and electron energy balances, respectively.  
Again, while the model is globally energy conserving, there can still be an error associated with the local energy conservation of each individual plasma species, as demonstrated by these results. 
Incidentally, recent work shows that implicit algorithms  nevertheless conserve \textit{total} energy (all particles plus fields) locally, at least in a Cartesian coordinate setting \cite{chacon2025local}.
On the other hand, while the model is not momentum conserving, it is evident that with the grid size, time step, and macroparticle count per cell used in this work, momentum error is small.

\subsection{Electron velocity distribution function}\label{sec:evdf}

\begin{figure}
    \centering
    \includegraphics[width=\textwidth]{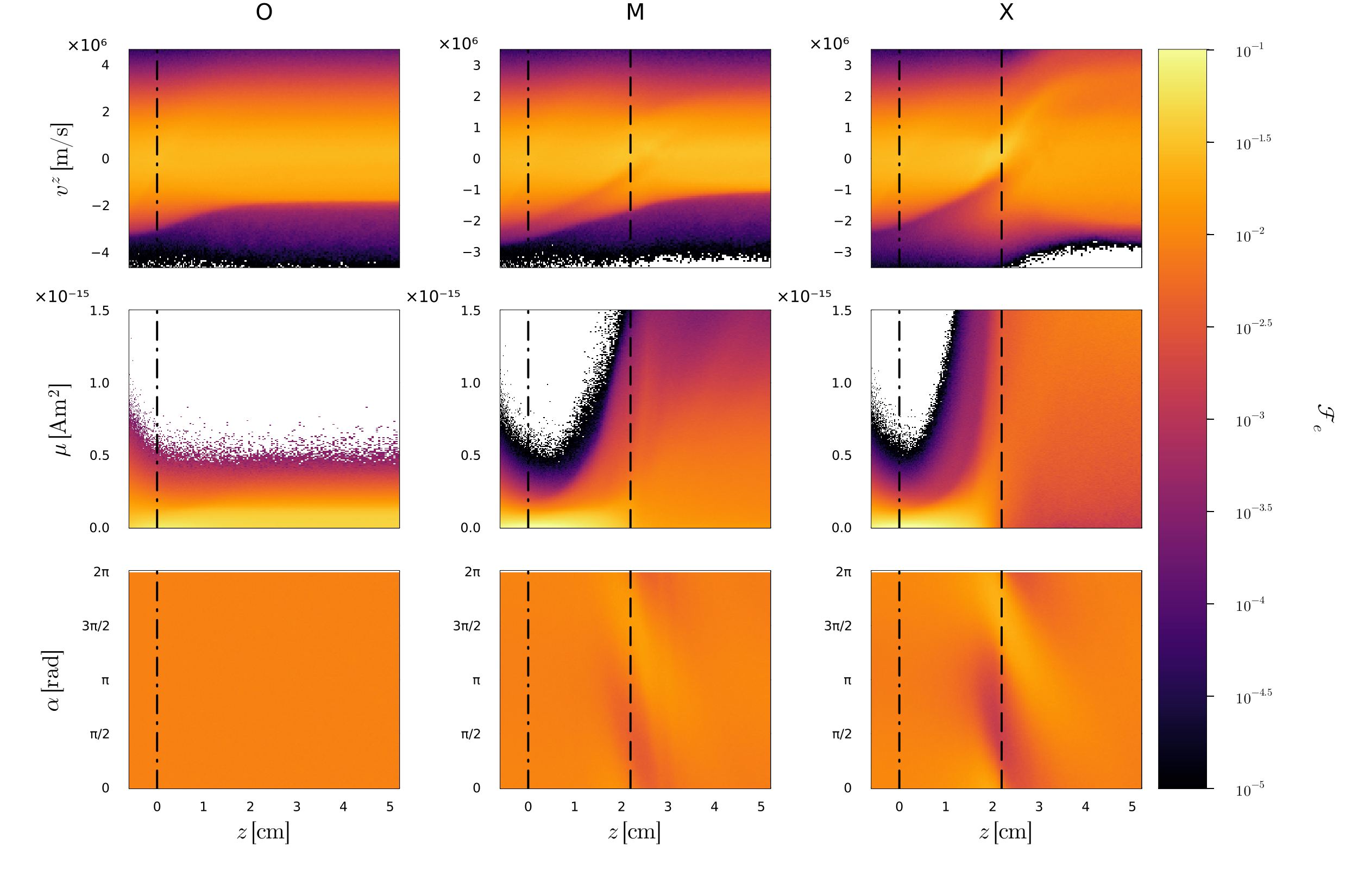}
    \caption{Normalized electron velocity distribution function $\fdist_e$ for the simulation cases O, M, X. Parallel velocity $v^z$ (first row), magnetic moment $\mu$ (second row) and gyrophase $\alpha$ (third row). The black vertical dash-dot and dash lines indicate the location of the throat of the MN ($T$) and the ECR ($R$).
    \rfone{The histograms at each $z$ are normalized such that their integral across $v^z$, $\mu$, or $\alpha$ equals 1. A total of
    128 bins were used in each velocity direction, and we average over the last 3000 time steps. This results in peak counts of macroparticles per bin of $6\cdot 10^5$, $12\cdot 10^5$, $3\cdot 10^5$ respectively upstream and 
    $3\cdot 10^5$, $1.2\cdot 10^5$, $0.9\cdot 10^5$ respectively downstream. In the gyrophase plots, the phase of the vector potential at $U$, $A_U$, is substracted prior to the averaging. }
    }
    \label{fig:evdf}
\end{figure}

\begin{figure}
    \centering
    \includegraphics[width=1\textwidth]{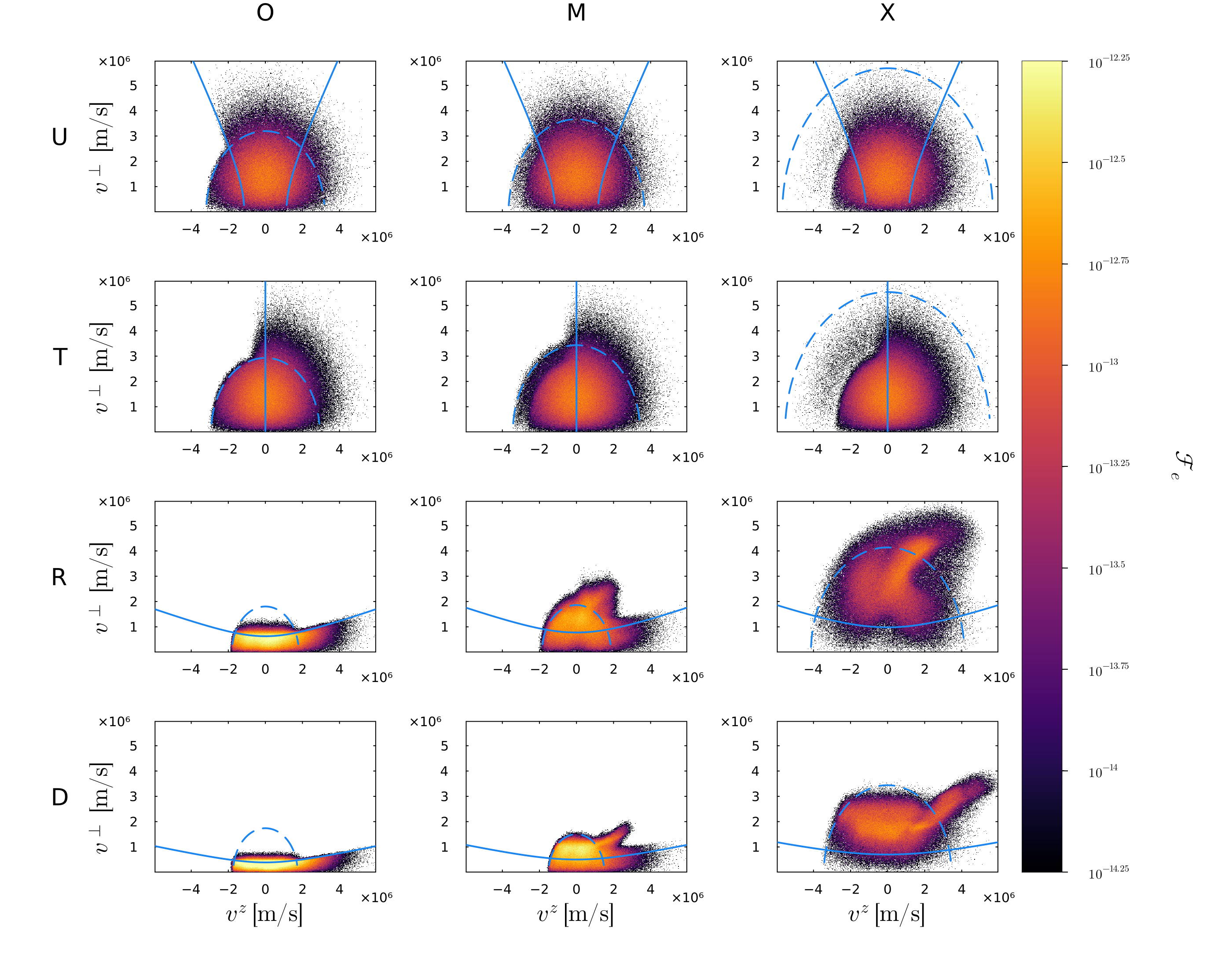}
    \caption{Normalized electron velocity distribution function $\fdist_e$ histograms in the $v^z$, $v^\perp$ plane at various locations in the MN: upstream (U), throat (T), resonance (R) and downstream (D).
    In each plot, the dashed line indicates the energy associated with $\phi(\zeta)-\phi_\infty$, which determines the phase space boundary of reflected/trapped and free electrons in an electrostatic setting; likewise, the solid line represents the  phase space boundary of reflected and trapped electrons.
    \rfone{Each histogram is normalized such that its integral across $v^z$ and $v^\perp$ equals 1. A total of
    400 bins were used in each velocity direction, and we average over the last 3000 time steps.
    This results in peak counts of macroparticles per bin of $3\cdot 10^4$ upstream and $75\cdot 10^4$ downstream.}
    }
    \label{fig:evdf2}
\end{figure}

To complete our analysis, we directly inspect the phase space of electrons of simulations O, M, and X in figures \ref{fig:evdf} and \ref{fig:evdf2}. While the former displays the evolution of $\fdist_e$ along $z$ in the early part of the domain, the latter presents snapshots of the $(v^z,v^\perp)$ plane at key locations in the MN.
\rfone{Depending on their mechanical energy and magnetic moment, electron trajectories are bound in $z$ electrostatically downstream, and by the magnetic mirror effect upstream.}
As in the work of Ahedo et al. \cite{ahed20a}, the curves shown in figure \ref{fig:evdf2} show the cutoffs associated with the potential at infinity $\phi_\infty$ and the magnetic mirror effect at the throat $T$, given by 
\begin{align}
(v^z)^2+(v^\perp)^2 &= -\frac{2e}{m_e}(\phi_\infty-\phi(\zeta))
\label{eq:circle}
\\
(v^z)^2+\left(1-\frac{B_{aT}}{B_a(\zeta)}\right)(v^\perp)^2 &= -\frac{2e}{m_e}(\phi_T-\phi(\zeta))
\label{eq:hyperbola}
\end{align}
In an electrostatic setting, electrons inside the circle of the first equation are confined electrons unable to escape to infinity.
The hyperbola of the second equation separates electrons that can traverse the MN throat from those that cannot. 

The electrostatic case O shows that $\fdist_e$ is cut off from below in the $v^z$ direction due to $\phi_\infty$. This causes anisotropy and asymmetry in the distribution, which results in an increasing $u^z_e$ that matches $u^z_i$ in spite of the decreasing potential $\phi$. 
As the expansion approaches infinity, the fraction of confined electrons gradually decreases, and the free electrons eventually become the only remaining population.
The magnetic mirror effect trades perpendicular electron energy for parallel electron energy as they move downstream in the divergent side: this explains why the normalized $\fdist_e$ essentially does not decrease in $v^z$ in figure \ref{fig:evdf}, and why the electrons seem to gradually displace clockwise toward the horizontal axis in figure \ref{fig:evdf2}.  

The view of $\fdist_e$ in terms of the magnetic moment in the second plot of figure \ref{fig:evdf} shows that $\mu$ remains essentially constant, with the exception of a small dip before the throat, consistent with the reflection of high-$\mu$ electrons in the convergent side.
This simulation case shows that the distribution is highly homogeneous in the gyrophase $\alpha$.

Interestingly, the population of \textit{doubly-trapped} electrons (the region between the circle and the hyperbola in the \rftwo{last} row of figure \ref{fig:evdf2}) is essentially empty. This region can only be filled in during the transient plume expansion or, mainly, thanks to collisionality (including numerical collisionality). While earlier works hypothesized full doubly-trapped regions \cite{mart15,meri21a}, the present result is in agreement with later numerical works \cite{sanc18b,zhou21a}, albeit this is known to be very sensitive to numerical diffusivity.
The emptiness of this region attests to the numerical accuracy of the simulation results.

Simulation M illustrates the effect in phase space of the heating by the electromagnetic wave. Most importantly, electrons gain magnetic moment $\mu$ as their perpendicular energy is enhanced by the absorption of the wave. Once the electrons are sufficiently downstream from the resonance $R$, $\mu$ continues to be conserved, and $\fdist_e$ becomes roughly flat in that direction.
The conversion of perpendicular energy to parallel energy due to the expanding $\vec B_a$ is responsible for the appearance of structures in the $v^z$ direction. The extra electron energy demands a stronger fall to $\phi_\infty$ to keep the MN current free, compared to the electrostatic case O, and this raises the cutoff $v^z$ in the first row of figure \ref{fig:evdf}. 
Interestingly, at the resonance the distribution function $\fdist_e$ develops an inhomogeneity in the gyrophase $\alpha$. This is the result of the majority of electrons aligning with the phase angle of the passing wave at $R$, which permits them to gain energy from it effectively. During the alignment process, however, out-of-phase electrons may lose some energy to the wave; this explains the observed behavior in figure \ref{fig:fields} near $R$, as suggested in Section \ref{sec:wave_fields}.

Figure \ref{fig:evdf2} shows that the wave provides an efficient mechanism to populate the doubly-trapped region of phase space, which remained empty in simulation O. This is the region between the drawn circle, within which electrons do not have enough energy to reach infinity downstream and will bounce back toward the source, and the upper side of the hyperbola, where electrons have a too high magnetic moment to pass through the throat.
In simulation M and X, this region is well populated.
Nevertheless, the resonant wave interaction is highly effective in transforming electrons among the three subpopulations (free, reflected, doubly-trapped), and indeed \rfone{the energy and magnetic moment bounds of
many of the doubly-trapped electrons 
allows them to undergo additional passes through the resonance}, which will modify their $v^\perp$ until they eventually are energetic enough to escape downstream or return to the source.

Those electrons, strongly energized by the wave with energies greater than the confining $\phi_\infty$, show up clearly in the last rows of fig. \ref{fig:evdf2}, sitting beyond the circle of equation \eqref{eq:circle}; these are free electrons that will not undergo another pass through the resonance. As the expansion proceeds downstream, the magnetic mirror effect displaces this population closer to the horizontal axis.

Interestingly, the circle no longer serves as a faithfull bound for the reflected electrons left from the throat for the same reason: there is a defect of reflected electrons at negative $v_z$ velocities, which is likely due to those electrons receiving an additional resonant pass as they go back to the source, increasing their $v^\perp$. Beyond the resonance, the circle becomes again the boundary between reflected and free electrons.

Finally, simulation X illustrates the fate of $\fdist_e$ under extreme wave heating. In this case, heating is so intense that the majority of electrons jump to high $\mu$ values as they cross the resonance $R$, as evidenced by the magnetic moment phase space of simulation X in figure \ref{fig:evdf}, where the most probable $\mu$ after the resonance is much greater than zero. \rfone{As depicted in the gyrophase $\alpha$ histogram of figure \ref{fig:evdf},  phase alignment of electrons is much more significant now}. The signature of the wave heating in the $v^z$ direction (again thanks to the mirror conversion) is also manifest: the large perpendicular energy is transformed into a peak in $\fdist_e$ in the $v_z$ direction, signaling a beam of hot electrons. This qualitatively different behavior can explain the changing trends in $T^\parallel_e$ and $q^\parallel_e$ at high wave power levels.

\section{Computational cost and numerical convergence}\label{sec:verification}

Verification, convergence and scaling studies for the simulation of a MN using the electrostatic version of this code were reported in \cite{jime24a}.
Furthermore, the electromagnetic version of the code has been verified against a Weibel instability reference case in \cite{jime24c} (not shown here). Below, 
we extend the performance analysis of the present formulation, beginning with the computational cost of the time-implicit Darwin-PIC algorithm. 
 
The resulting computational model compares positively to a traditional time-explicit PIC-Maxwell in terms of execution time. 
Assuming the same number of macroparticles per cell $N_p$ for an explicit and an implicit code, a back-of-the-envelope cost estimate like the one in \cite{chen14b} suggest the following scaling of the CPU wall time: 
\begin{equation}
    \frac{C P U_{exp}}{C P U_{imp}}
    \sim
    \left(
    \frac{\Delta z_{imp}}{\Delta z_{exp}}\right)^d
    \left(
    \frac{\Delta t_{imp}}{\Delta t_{exp}}
    \right)
    \frac{C_{exp}}{C_{imp}}
        ,
    \label{eq:costestimate}
\end{equation}
where $\Delta z_{imp}$, $\Delta z_{exp}$, $\Delta t_{imp}$, $\Delta t_{exp}$  represents the characteristic cell size and timestep in each code, $d$ is the physical space dimensionality of the problem, and $C_{exp}/C_{imp}$ is the ratio of computational complexities of each solver. If the implicit code uses $N_{FE}$ JFNK iterations per time step and a typical substep $\Delta\tau$,
$C_{imp}\sim C_{exp} N_{FE}\Delta t_{imp}/\Delta \tau $ (assuming the cost of moving the macroparticles in a substep $\Delta\tau$ is comparable to the cost of the explicit solver). This makes the cost estimate \eqref{eq:costestimate} independent of the timestep $\Delta t_{imp}$, which instead depends on the substep $\Delta\tau$.

A first cause of the computational cost savings of the implicit model is its capability to employ larger cell sizes. Whereas in an explicit code it must be a fraction of the local Debye length (typically, $\Delta z_{exp} \simeq 0.1$--$0.2 \lambda_D^*$) to avoid numerical instabilities, in the implicit code we can pick a cell size comparable to the gradient size of the macroscopic quantities. There is, however, a tradeoff: a larger $\Delta z_{imp}$ typically requires a greater number of nonlinear iterations $N_{FE}$. Nevertheless, for our choice of $\Delta z = 20\lambda_D^*$, the first factor in \eqref{eq:costestimate} is about 100 for $d=1$.
Secondly, our substep $\Delta \tau$ is significantly greater than what the speed of light CFL condition would dictate:  
for cells of $\Delta z_{exp} =0.2 \lambda_D^*$, $\Delta t_{exp}<0.016$ ps is required. 
For our $\Delta z_{imp}=20\lambda_D^*$, this would be $1.57$ ps or less, but instead we get to operate with a typical value of $\Delta \tau \simeq 60$ ps.
This represents a large saving factor greater than 3000.
Finally, the nonlinear solution of the potentials requires typically $14$ particle push iterations
(the product of 1--4 GMRES iterations for 3--6 Newton iterations).
with a relative tolerance of $10^{-4}$.
This is moderate and is clearly offset by the gains listed above for our 1D simulation; the advantage of the implicit approach could be even greater in higher dimensions.
All in all, the wall time of each $5$ $\mu$s simulation with our current (unoptimized) implementation is about $114$ hours on a single machine with an Intel Xeon Gold 6230 using 20 cores. 

To complement the main simulations in this study, we present a convergence study for the intermediate case M. Additional cases are considered that degrade the numerical parameters of the nominal M: case M$(2\Delta t)$ duplicates the time step, M$(N_p/2)$ roughly halves the number of macroparticles per cell, and M$(N_g/2)$ halves the number of grid cells. We also include a fourth extra case M(ext) that extends the domain size by $35\%$, while keeping approximately the same cell size map $\ff$ and macroparticle per cell count as the nominal M.

\begin{figure}
\includegraphics[width=1\textwidth]{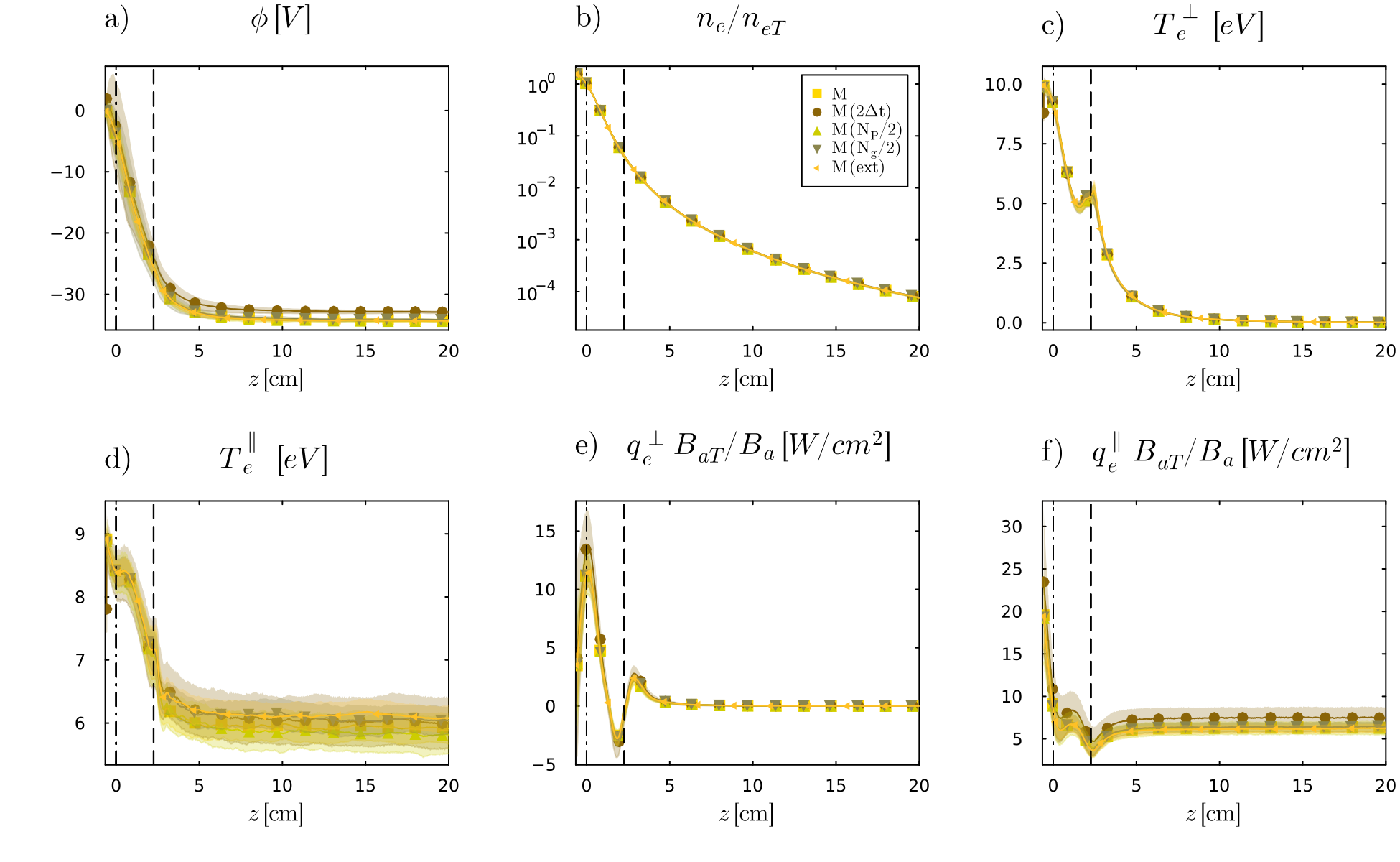}
\centering
\caption{Steady-state solution of (a) $\phi$, (b) $n_i$, (c) $T^\perp_{e}$, (d) $T^\parallel_{e}$, (e) $q^\perp_{e}$, and (f) $q^\parallel_{e}$ for the nominal M case and the additional simulation cases of the convergence study. Solid lines and shaded areas indicate the mean and the standard deviation, computed over the last 1500 time steps. See legend in plot (b) for the meaning of each line. The black vertical dash-dot and dash lines indicate the location of the throat of the MN ($T$) and the ECR ($R$).}
\label{fig:convergence}
\end{figure}

The simulation results of these extra cases are displayed in figure \ref{fig:convergence} and a summary of their relative errors is presented in table \ref{tab:convergencescaling}, showing good convergence for all variables. The only noteworthy exceptions are $\phi$, $T^\parallel_e$, and $q^\parallel_e$ which do show some minor yet visible differences with the M simulation, especially for simulation M($2\Delta t$). 
\rfone{
In addition we note the small variations in the magnitude of $\vec A$ across the verification simulations, in particular the extended domain case M(ext) (shown only in table \ref{tab:convergencescaling}). This allows us to argue that  the simplified downstream boundary condition for $\vec A$, stated in Section \ref{sec:fields}, does not induce any relevant domain size-dependent wave reflections.}

\begin{table}
\centering
\begin{tabular}{c|c|c|c|c|c} 
  Case& $err_\phi$& $err_{T_e^{\parallel}}$&$err_{q_e^{\parallel}}$ &$err_{A}$&$t$ [h]\\\hline
  $\mathrm{M}$& -  & - &- &-&$114 $\\ 
  $\mathrm{M(2\Delta t)}$& $1.46  \cdot 10^{-2} $& $4.59 \cdot 10^{-4} $&$ 2.75 \cdot 10^{-5}$ &$ 6.52 \cdot 10^{-7}$&$39 $\\ 
$\mathrm{M(N_p/2)}$& $3.12  \cdot 10^{-4} $& $2.12  \cdot 10^{-4} $&$ 1.86 \cdot 10^{-7}$ &$ 4.68 \cdot 10^{-7}$&$47$\\ 
$\mathrm{M(N_g / 2)}$& $1.24  \cdot 10^{-4} $& $3.25  \cdot 10^{-4} $&$ 1.81 \cdot 10^{-7}$ &$ 3.16 \cdot 10^{-7}$&$43$\\ 
$\mathrm{M(ext)}$& $1.55  \cdot 10^{-4} $& $3.18  \cdot 10^{-4} $&$ 4.35 \cdot 10^{-7}$ &$ 2.84 \cdot 10^{-7}$&$175$\\   \end{tabular}
\caption{
Normalized, integrated square difference of the mean values with respect to the reference one of $\phi$,  $T^\parallel_{e}$,  $q^\parallel_{e}B_{aT}/B_a$, \rfone{and $A$}
for the simulation cases of the convergence study (computed as
{${\int (\hat Q - \hat Q_R)^2 \dd z}/(z_D - z_U)$}, where $\hat Q$ is one of those profiles and $\hat Q_R$ the reference profile for simulation M, both of them made dimensionless as required with $e$, $n_i^*$, $T_e^*$, $m_e$). For simulation M(ext), only the part of the domain in common with simulation M is used for the error computation. Such errors in $n_e$ are of order $ 10^{-7}$, in $T_e^\perp$ and $q_e^\perp$ of $ 10^{-5}$ or less.
All simulations were run on an Intel Xeon Gold 6230.  }
\label{tab:convergencescaling}
\end{table}

The simulation time drops to $39$--$47$ h for  M($2\Delta t$), M($N_p/2$) and M($N_g/2$). 
Given the good accuracy of these degraded simulation cases, 
this suggests that our choice of numerical parameters for the main simulations is conservative and can be relaxed.
Indeed, while the practical limits of $\Delta t$, $N_p$, and $N_g$ have not been explored in the present work,
it is evident that there is potential for further savings in the computational cost before significant divergences with the nominal simulation develop.
In addition, while our preliminary implementation employs some basic parallelization, we note that it is not optimized, as it is intended to demonstrate the ability of the algorithm to solve relevant plasma thruster problems. 
All this hints at the possibility of much faster simulations (by optimizing the code and correctly selecting the numerical parameters) for most applications.

\rfone{For completeness, we also performed a similar convergence study for the simulation with highest wave power, X, which displayed the largest standard deviation in figure \ref{fig:steady}.
In this case, while we do observe convergence, we noticed a larger spread across all verification simulations in the mean of $\phi$ (about 5 V over $69$ V), $T^\parallel_e$ (about 2.5 eV over $20$ eV) and $q^\parallel_e B_{aT}/B_a$ (about 5 W/cm$^2$ over 24 W/cm$^2$).
}

\section{Conclusion}\label{sec:conclu}

The collisionless kinetic response of a magnetized plasma expanding in a MN under the effects of a propagating electromagnetic wave and its absorption at the electron cyclotron resonance have been investigated, thanks to 
a novel quasi-1D, fully-implicit, energy- and charge-conserving PIC model. 

Our results have shown that an incident electromagnetic wave can substantially modify the physics of the plasma expansion in a MN with respect to the electrostatic case. The most remarkable aspects are the localized rise of $T^\perp_e$ at the resonance (reminiscent of previous experimental measurements and simulations of HPTs with this feature in the MN), the consequent larger fall of $\phi$ in the MN, and the non-monotonic behavior of $T^\parallel_e$ and $q^\parallel_e$, which hints at a complex interplay between the wave and the mean and thermal electron energy downstream from the resonance.

We have found that the self-consistent electromagnetic fields \rfone{can be reproduced with a} properly-tuned cold plasma dielectric tensor model. 
This was done by defining an `effective' collisionality to ensure a good fit with the kinetic results.
Ideally, this approach should enable the calibration of faster, simpler models that, although not kinetically self-consistent, may enable sufficient insight to part of the physics of electrodeless plasma thrusters. However, it is evident that such surrogate models require empirical tuning and miss key aspects of the wave-plasma interaction physics.

The acceleration of ions in the MN is enabled by the thermal energy. In electron-driven MNs, this is mostly in the electron population. The magnetic mirror effect converts the perpendicular energy into parallel energy. Parallel electron energy is then transformed into parallel ion energy owing to the mediation of the electrostatic field. 
These are the dominant terms in the momentum and energy balances of ions and electrons.
From a momentum-balance viewpoint, the reaction to the magnetic mirror force is felt at the magnets or solenoids that create $\vec B_a$: this is the magnetic thrust, which together with the initial flow of plasma momentum equals the total thrust delivered by the system. From an energy-balance viewpoint, the expansion is fueled by the electron thermal energy.

The addition of the wave enhances this thermal energy at the resonance, and thus results in a larger potential fall, a larger ion acceleration, and a greater MN thrust. Interestingly, the thrust efficiency of the system (taking into account all power inputs) can increase slightly with respect to the electrostatic case under small values of the wave power; however, 
it diminishes as the wave amplitude is increased beyond a certain limit value, when a substantial part of the wave power ends up as frozen losses in the parallel electron heat flux and convective thermal flux. \rfone{The threshold wave power seems to depend on the value of the ion-to-electron mass ratio, $m_i/m_e$.}

The separation between reflected, doubly-trapped, and free electrons blurs in the presence of the wave, as it provides an efficient mechanism to transfer electrons among these three populations.
In particular, the wave enables filling up the doubly-trapped region, otherwise essentially empty (in the absence of collisions). Moreover, it also populates the region of downstream phase space composed of high-$\mu$ electrons, which is inaccessible in the electrostatic case. These electrons are also free electrons, and their existence can skew their distribution, especially at high wave powers.

One of the key advantages of the time-implicit formulation used, compared to traditional time-explicit PIC codes, is that we can reliably increase our cell size and time step over the local Debye length and plasma time, which provides a clear avenue to accelerate kinetic plasma simulations.
The use of the Darwin electromagnetic model is highly appropriate in dense plasmas and removes the stiffness of the speed of light.
\rfone{Error figures in the convergence analysis are remarkable, and estimated computational times suggest this algorithm could be} competitive with respect to traditional, momentum-conserving, explicit PIC codes (especially if they implement the Maxwell electromagnetic model), even when we have indulged in a relatively large macroparticle per cell count to obtain low noise results.
A careful assessment requires an actual benchmark, not afforded here.
This class of PIC algorithms is \rfone{an interesting candidate} for the design, characterization, and optimization of electric propulsion devices, including those where electromagnetic fields play an important role, such as electrodeless plasma thrusters. The algorithm speed and its conservation properties make it an interesting \rfone{building block in the pursuit toward} accurate, quantitatively-predictive device models that \rfone{might one day} offload part of the responsibility of experimental test campaigns in the long and expensive thruster lifetime characterization, by numerically accelerating them. 

\appendix
\section{Differential geometry}
\label{sec:diffgeometry}

\begin{figure}
    \centering
    \includegraphics[width=0.5\linewidth]{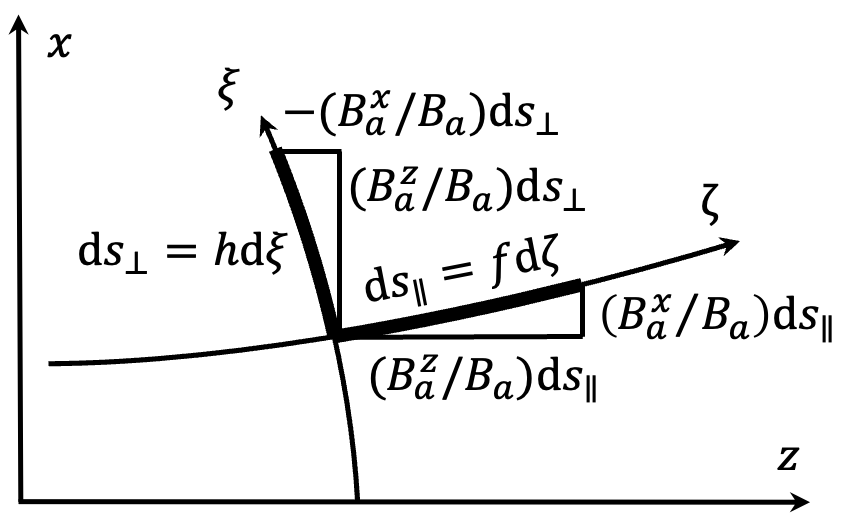}
    \caption{Sketch of the curvilinear coordinates $(\xi,\zeta)$ in the $(x,z)$ plane. A similar definition of $\eta$ follows in the $(y,z)$ plane.}
    \label{fig:sketch_coord}
\end{figure}

The null divergence of the applied magnetic field $\vec B_a$ and its axisymmetry give, at the axis: 
\begin{align}
\pd_x B_a^x =\pd_y B_a^y = -\frac{1}{2}\pd_z B_a^z
\end{align}
We consider a thin magnetic tube of interest about the axis, $x,y\sim \ell \sim \varepsilon L$, where $\ell$ is the largest Larmor radius of interest and $L$ is the axial gradient length.
In this tube, $B_a^z(x, y, z) = B_a^z(0,0,z) + O(\varepsilon^2)$ holds, since $B_a^z$ is an even function of $x$ and $y$.
To avoid overcomplicating our notation, we refer to the magnetic field at the axis simply as $B_a^z(z)$.
Then, $B_a^x(x, y, z) = -(x/2)\pd_z B_a^z(z) + O(\varepsilon^3)$, and similarly for $B_a^y$.
The magnetic field at a point $(x, y, z)$ in this tube is therefore:
\begin{align}
\vec B_a (x,y,z) =
-\frac{x}{2}\pd_z B_a^z(z) \vec 1_x
-\frac{y}{2}\pd_z B_a^z(z) \vec 1_y
+ B_a^z(z)\vec 1_z+O(\varepsilon^2).
\end{align}
Define a stream-wise coordinate $\zeta$ on the magnetic lines of our tube, such that the arc length along it is $\dd s_\parallel = \ff \dd \zeta$. Further, define an orthogonal coordinate $\xi$ to it, such that the arc length along a line perpendicular to the streamline is $\dd s_\perp = \hh \dd \xi$ in the $(x,z)$ plane. 
Figure \ref{fig:sketch_coord} represents the $(x,z)$ plane and the definition of the curvilinear coordinates $(\xi,\zeta)$. 
For completeness, we introduce a third orthogonal coordinate $\eta$, analogous to $\xi$ but in the $(y,z)$ plane.

The quantities $\ff$ and $\hh$ are $O(1)$ functions of $\zeta$; $\ff(\zeta)$ allows to define non-uniform grids of the domain ($\ff=1$ for a uniform grid), while function $\hh(\zeta)$ is fixed as
$\hh(\zeta)=\sqrt{B_{aT}^z/B_a^z(z(\zeta))}$, which represents the normalized radius of the magnetic tube. Observe that $\hh=1$ at the throat $T$.
Together, they constitute the \textit{exact} (by ansatz) Lamé factors of the orthogonal transformation 
$(\xi,\eta, \zeta)\to(x,y,z)$, whose metric tensor is
\begin{align}
G &=  
\left[
\begin{array}{ccc}
\hh^2 & 0 & 0 \\
0 & \hh^2 & 0 \\
0 & 0 & \ff^2 
\end{array}
\right].
\label{eq:metrictensor}
\end{align}
The associated volume factor is $\sqrt{g} = \sqrt{\det(G)} = \hh^2f$. 

Denoting derivatives with respect to $\zeta$ with a prime $(')$, the slope of the magnetic lines is determined by:
\begin{align}
\hh' \equiv \frac{\dd \hh}{\dd \zeta} =  
-\frac{\hh}{2} \pd_\zeta \ln B^z_a.
\end{align}
The approximate Jacobian of the transformation 
and its inverse are:
\begin{align} 
J
&=\left[
\begin{array}{ccc}
\hh & 0 & \xi \hh' \\
0 & \hh & \eta \hh' \\
-\xi \hh \hh'/\ff & -\eta \hh \hh'/\ff & \ff
\end{array}
\right];
&
J^{-1}
&= \left[
\begin{array}{ccc}
1/\hh & 0 & -\xi \hh'/(\hh \ff) \\
0 & 1/\hh & -\eta \hh'/(\hh \ff) \\
\xi \hh'/\ff^2 & \eta \hh'/\ff^2 & 1/\ff
\end{array}
\right],
\label{eq:jacobian}
\end{align}
both with an error of $O(\varepsilon^2)$.
Integrating along the axis on $\zeta$ and then on a perpendicular $\xi,\eta$ surface, we find the approximate expressions for the transformation itself, 
\begin{align}
x &= \hh\xi + O(\varepsilon^3),
&
y &= \hh\eta + O(\varepsilon^3),
&
z &= \int_0^\zeta \ff\dd \zeta -\frac{\hh \hh'}{2\ff}(\xi^2 + \eta^2) + O(\varepsilon^3).
\label{eq:coordinate_transform}
\end{align}
Observe that $J$, $J^{-1}$, and these transformation expressions become \textit{exact} at the axis ($\xi,\eta =0$).

Incidentally, we note that the MN need not be \textit{paraxial}, i.e., we do not require the stronger ordering $\ell\sim \varepsilon R \sim \varepsilon^2 L\sim \varepsilon^3\lambda_c$, where $R$ is the radial gradient length, distinguished from the axial gradient length $L$. Adding this additional requirement would double the order of accuracy of most expressions in the model, but this is not strictly necessary for the model to hold.

The local vector basis $\{\ee_\xi,\ee_\eta, \ee_\zeta\}$ associated with the $\{\xi,\eta,\zeta\}$ coordinates is given by the columns of $J$; likewise, the local covector basis $\{\cee^\xi, \cee^\eta, \cee^\zeta\}$ is given by the rows of $J^{-1}$. Hence, with an error $O(\varepsilon^2)$ outside of the axis,
\begin{align*}
\ee_\xi &\simeq \hh\vec 1_x - \frac{\hh \hh'}{\ff}\xi \vec 1_z; 
&
\ee_\eta &\simeq \hh\vec 1_y - \frac{\hh \hh'}{\ff}\eta \vec 1_z; 
&
\ee_\zeta &\simeq \xi \hh'\vec 1_x + \eta \hh'\vec 1_y +\ff\vec 1_z;
\\
\cee^\xi &\simeq \frac{1}{\hh}\cov 1^x -\xi\frac{\hh'}{\hh \ff}\cov 1^z; 
&
\cee^\eta &\simeq \frac{1}{\hh}\cov 1^y
-\eta\frac{\hh'}{\hh \ff}\cov 1^z; 
&
\cee^\zeta &\simeq \frac{\hh'}{\ff^2}\xi \cov 1^x + \frac{\hh'}{\ff^2}\eta \cov 1^y + \frac{1}{\ff}\cov 1^z;
\end{align*}
The two bases introduced above are the dual of each other, since $\cee^{i}\cdot\ee_j = \delta^i_j$. In the present work, we abuse the dot notation ($\cdot$) to indicate the adequate contraction in each case: either the inner product of two vectors with the metric tensor, or the application of a covector on a vector (or vice versa).

Any tangent vector $\vec a$ and a covector $\cov \alpha$ can be expressed in terms of these local bases as:
\begin{align}
\vec a &= a^\xi\ee_\xi + a^\eta\ee_\eta + a^\zeta\ee_\zeta,
\\
\cov \alpha &= \alpha_\xi \cee^\xi + \alpha_\eta \cee^\eta + \alpha_\zeta \cee^\zeta .
\end{align}
Again abusing language, we  speak of the covariant components of a vector $\vec a$ as those defined as $a_i =g_{ij}a^j$, and the contravariant components of a  covector $\cov \alpha$, as $\alpha^i=g^{ij}\alpha_j$.

Given a vector in this form (or respectively, a covector), its Cartesian components $a^x,a^y,a^z$ (respectively, $\alpha_x,\alpha_y,\alpha_z$) in the canonical global basis $\{\vec 1_x, \vec 1_y, \vec 1_z\}$ (respectively, $\{\cov 1^x, \cov 1^y, \cov 1^z\}$) are:
\begin{align}
a^x  &= \hh a^\xi +\xi \hh'a^\zeta; &a^y  &= ha^\eta+\eta \hh'a^\zeta; 
&
a^z &= - \xi \frac{\hh \hh'}{\ff} a^\xi  - \eta \frac{\hh \hh'}{\ff}a^\eta + fa^\zeta,
\\
\alpha_x &= \frac{\alpha_\xi}{\hh} + \xi \frac{\hh'}{\ff^2} \alpha_\zeta; 
&
\alpha_y &= \frac{\alpha_\eta}{\hh}  + \eta \frac{\hh'}{\ff^2} \alpha_\zeta; 
&
\alpha_z &= -\xi \frac{\hh'}{\hh \ff}\alpha_\xi - \eta\frac{\hh'}{\hh \ff}\alpha_\eta + \frac{\alpha_\zeta}{\ff}.
\end{align}
Note that in the canonical vector basis there is no value distinction between contravariant and covariant components.

The most general functional structure of a scalar field $\phi$ and a vector field $\vec a$  which satisfy the symmetry conditions of the problem,
exact at the axis and with error $O(\varepsilon^4)$ away from it,
is as follows: 
\begin{align}
\phi\simeq\Phi(\zeta)+\frac{\kappa_\phi(\zeta)}{4}(\xi^2+\eta^2), 
\\
a^{q^l} \simeq A^{q^l}(\zeta)
+
\frac{\kappa_{q^l}(\zeta)}{4}(\xi^2+\eta^2), 
\end{align}
where $\Phi(\zeta), \kappa_\phi(\zeta),A^{q^l}(\zeta), \kappa_{q^l}(\zeta)$ (with $q^l=\xi,\eta,\zeta$) are functions of $\zeta$.
As explained in the main text, the electrostatic and magnetic potentials $\phi$ and $\vec A$ in this work do not have the extra curvature terms $\kappa_\phi$, $\kappa_{q^l}$, and consequently, their expressions simplify further.
For such fields, the gradient, curl, and divergence, exact at the axis and with error $O(\varepsilon^3)$ elsewhere, are:
\begin{align}
\nabla\phi &\simeq  
\frac{1}{\ff^2} \pd_\zeta (\Phi) \ee_\zeta.
\\
\nabla\times\vec a &\simeq
\frac{1}{\hh^2 \ff}\left(
- \pd_\zeta( \hh^2 A^\eta)\ee_\xi
+
\pd_\zeta (\hh^2 A^\xi)\ee_\eta
\right)
\\
\nabla\cdot\vec a &\simeq  \frac{1}{\hh^2 \ff}\frac{\pd }{\pd\zeta}\left(\hh^2 \ff A^\zeta\right).
\end{align}

Finally, the Laplacian of $\phi$ and the curl-curl of $\vec a$, exact at the axis
and with error $O(\varepsilon^2)$ elsewhere, are:
\begin{align}
\nabla^2\phi &=  
\frac{1}{\hh^2f}\pd_\zeta\left(\frac{\hh^2}{\ff}\pd_\zeta \Phi\right),
\\
\nabla\times(\nabla\times\vec a) &= 
-\frac{1}{\hh^2f} \pd_\zeta\left(\frac{1}{\ff}\pd_\zeta( \hh^2 A^\xi)\right)\ee_\xi 
-\frac{1}{\hh^2f} \pd_\zeta\left(\frac{1}{\ff}\pd_\zeta( \hh^2 A^\eta) \right)\ee_\eta .
\end{align}
Note that, at the axis, $A^\xi=A^x/\hh$, $A^\eta=A^y/\hh$, $\vec e_\xi = \hh\vec 1_x$ and $\vec e_\eta = \hh \vec 1_y$, which helps relate the local basis vector components back to the canonical vector basis components.

As a final marginal note, we observe that while in this work we have opted for the computational coordinates $\xi,\eta,\zeta$, we note that a different parametrization in $\rho=\sqrt{\xi^2+\eta^2},\alpha=\arctan(\eta/\xi),\zeta$ is possible. 
When combined with cylindrical velocity coordinates $v^z,v^\perp$, it offers some advantages over our present choice.
In those cylindrical coordinates, $\rho$ varies slowly and $\alpha$ varies almost linearly, which has clear merits for numerical integration with longer timesteps, presently constrained by the gyrofrequency of electrons due to the need to integrate the quickly varying $v^x$ and $v^y$ of the particles.
Nevertheless, such parametrization is harder to extend to higher dimensions, when the magnetic streamlines can have any spatial orientation. As our code is intended to serve as a proof-of-concept before attempting higher dimensions, this determined our choice of parametrization.
 
\section{Shape functions}\label{sec:shapefunctions}

We denote as $S_i^m(\zeta)$ the order-$m$ B-spline $B^m(\zeta; \zeta_i)$ centered at $\zeta_i$. In particular, we have
\begin{align}
S_i^0(\zeta) = B^0(\zeta; \zeta_i) &= 
\begin{cases}
1 & \mbox{if } |\zeta-\zeta_i| < 1/2 
\\
0 & \mbox{otherwise;}
\end{cases}
\\
S_i^1(\zeta) = B^1(\zeta; \zeta_i) &= 
\begin{cases}
1  - |\zeta-\zeta_i|  & \mbox{if } |\zeta-\zeta_i| < 1
\\
0 & \mbox{otherwise;}
\end{cases}
\\
S_i^2(\zeta) = B^2(\zeta; \zeta_i) &=
\begin{cases}
3/4  - |\zeta-\zeta_i|^2  & \mbox{if } |\zeta-\zeta_i| < 1/2 
\\
1/2(3/2  - |\zeta-\zeta_i|)^2& \mbox{if } 1/2 \leq |\zeta-\zeta_i| < 3/2
\\
0 & \mbox{otherwise.}
\end{cases}
\end{align}
Observe that the support of the shape function $S^m_i$ has a width 
$\Delta \zeta = m+1$, centered at $\zeta=i$.
The definition of the shape functions at the half-integer grid positions, $S_{i+1/2}^m(\zeta)$, is analogous.

Key properties of shape functions are:
\begin{align}
S_i^m(\zeta) &= \int_{-\infty}^\infty S_i^{m-1}(\zeta') B^0(\zeta'; \zeta) \dd \zeta';
\\
\frac{\dd}{\dd \zeta} S_i^m(\zeta) &= S_{i-1/2}^{m-1}(\zeta) - S_{i+1/2}^{m-1}(\zeta);
\\
\int_{-\infty}^\infty S^m_i(\zeta') \dd \zeta' &= 1;
\\
\int_{-\infty}^\zeta S_i^m(\zeta')\dd \zeta' &= \sum_{j=i}^\infty S_{j+1/2}^{m+1}(\zeta) 
\\
\sum_i S^m_i(\zeta) &= 1 \quad \forall \zeta;
\\
\sum_j S^m_i(\zeta_j) &= 1 \quad \forall i;
\end{align}

\section*{Acknowledgments}

This project has received funding from the European Research Council (ERC) under the Horizon 2020 research and innovation program of the European Union (project ERC-StG ZARATHUSTRA, grant agreement No. 950466). 
Additional funding came from the R\&D project PID2023-150052OB-I00 (ADAPT) funded by MICIU/AEI/10.13039/501100011033 and by ERDF, EU. Luis Chacón's contributions were performed by Los Alamos National Laboratory under the auspices of the Office of Advanced Scientific Computing Research of the U.S.\ Department of Energy under contract 89233218CNA000001.

The Department of Aerospace Engineering at Universidad Carlos III de Madrid has been accredited as a Unit of Excellence “María de Maeztu”, a distinction awarded by the Spanish State Research Agency for the period 01/01/2026-31/12/2031.

\bibliographystyle{ieeetr}
\bibliography{bibtex/ep2,bibtex/others,bibtex/references,bibtex/luis}

@STRING{adv = {Adri\'an Dom\'{i}nguez-V\'{a}zquez}}

@STRING{amc = {Alberto Mar\'{i}n-Cebri\'{a}n}}

@STRING{bt  = {Bin Tian}}

@STRING{ea  = {Eduardo Ahedo}}

@STRING{ebb = {Enrique Bello-Ben\'itez}}

@STRING{hu  = {Hodei Urrutxua}}

@STRING{jn  = {Jaume Navarro-Cavall\'{e}}}

@STRING{jz  = {Jiewei Zhou}}

@STRING{jr  = {Jes\'{u}s Ramos}}

@STRING{ml  = {Min Li}}

@STRING{mi  = {Marco Riccardo Inchingolo}}

@STRING{mm  = {Mario Merino}}

@STRING{pj  = {Pedro Jim\'enez}}

@STRING{and = { and }}

@STRING{PoP = {Physics of Plasmas}}

@STRING{psst = {Plasma Sources Science and Technology}}

@STRING{gec76 = {$76^{th}$ } # gec}

@STRING{iepc38 = {$38^{th}$ } # iepc}

@STRING{AIAA = {American Institute of Aeronautics and Astronautics}}

@STRING{iop = {IOP Publishing}}

@STRING{erps = {Electric Rocket Propulsion Society}}

@article{madd25a,
    author = {Davide Maddaloni and Borja Bayón-Buján and Jaume Navarro-Cavallé and Mario Merino},
    title = {Low-frequency oscillations in the magnetic nozzle of a Helicon Plasma Thruster},
    journal = {Plasma Sources Scinece and Technology},
    publisher={IOP Publishing},     
    url = {https://iopscience.iop.org/article/10.1088/1361-6595/adc40d},
    year={2025},
volume = {34},
number = {4}
}

@article{jime24a,
    title = {An implicit, conservative electrostatic particle-in-cell algorithm for paraxial magnetic nozzles},
    journal = {Journal of Computational Physics},
    volume = {502},
    pages = {112826},
    year = {2024},
    issn = {0021-9991},
    doi = {https://doi.org/10.1016/j.jcp.2024.112826}, 
    author = {Pedro Jiménez and Luis Chacón and Mario Merino},
    publisher={Elsevier}
}

@article{jime23a,
title={Analysis of a cusped helicon plasma thruster discharge},
  author={Jim{\'e}nez, Pedro and Zhou, Jiewei and Navarro-Cavall{\'e}, Jaume and Fajardo, Pablo and Merino, Mario and Ahedo, Eduardo},
  journal={Plasma Sources Science and Technology},
  doi = {10.1088/1361-6595/ad01da},
  year={2023},
  publisher = {IOP Publishing},
  volume = {32},
  number = {10},
  pages = {105013},
}

@article{svil23,
author={\'A. S\'anchez-Villar and F. Boni and V. D\'esangles and J. Jarrige and D. Packan and E. Ahedo and M. Merino},
title={Comparison of a hybrid model and experimental measurements for a dielectric-coated coaxial {ECR} thruster},
journal = psst,
doi = {10.1088/1361-6595/acb00c},
year = {2023},
publisher = {IOP Publishing},
volume = {32},
number = {1},
pages = {014002},
}

@article{cich22a,
  AUTHOR={Cichocki, Filippo and Navarro-Cavall\'e, Jaume and Modesti, Alberto and Ram\'irez V\'azquez, Gonzalo},   
  TITLE={Magnetic Nozzle and RPA Simulations vs. Experiments for a Helicon Plasma Thruster Plume},      
  JOURNAL={Frontiers in Physics},      
  VOLUME={10},      
  YEAR={2022},      
  DOI={10.3389/fphy.2022.876684},      
  ISSN={2296-424X} 
}

@article{meri21a,
    year = 2021,  
    author = {Mario Merino and Judit Nuez and Eduardo Ahedo},
    title = {Fluid-kinetic model of a propulsive magnetic nozzle},
    journal = psst,
    volume = {30},
    number = {11},
    pages = {115006},
    publisher = {{IOP} Publishing},
    ISSN = {1089-7674},
    doi = {10.1088/1361-6595/ac2a0b}
}

@Article{zhou21a,
  Title                    = {Time-dependent expansion of a weakly-collisional plasma beam in a paraxial magnetic nozzle},
  Author                   = {Zhou, Jiewei and S{\'a}nchez-Arriaga, Gonzalo and Ahedo, Eduardo},
  Journal                  = psst,
  doi                      = {10.1088/1361-6595/abeff3},
  Year                     = 2021,
  Number                   = {4},
  Pages                    = {045009},
  Volume                   = {30},
  Publisher                = iop
}

@article{svil21a,
  doi = {10.1088/1361-6595/abde20},
  author = {A. S\'anchez-Villar and J. Zhou and M. Merino and E. Ahedo},
  title = {Coupled plasma transport and electromagnetic wave simulation of an {ECR} thruster},
  journal = psst,
  volume = {30},
  number = {4},
  pages = {045005},
  ISSN = {1089-7674},
  year = 2021,
  publisher = {{IOP} Publishing}
}

@article{ahed20a,
	doi = {10.1088/1361-6595/ab7855}, 
	year = 2020, 
	publisher = {{IOP} Publishing},
	volume = {29},
	number = {4},
	ISSN = {1089-7674},
	pages = {045017},
	author = {Eduardo Ahedo and Sara Correyero and Jaume Navarro and
	Mario Merino},
	title = {Macroscopic and parametric study of a kinetic plasma expansion in a paraxial magnetic nozzle},
	journal = {Plasma Sources Science and Technology}
}

@article{li19a,
 title={On electron boundary conditions in {PIC} plasma thruster plume simulations},
 author= ml # and # mm # and # ea # and # {Haibin Tang},
 journal= psst,
 volume={28},
 number={03},
 pages={034004},
 year={2019},
 doi = {10.1088/1361-6595/ab0949}, 
 publisher= iop,
 ISSN = {0963-0252} 
}

@article{tian18a,
 author= bt # and # mm # and # ea,
 title={Two-dimensional plasma-wave interaction in an helicon plasma thruster with magnetic nozzle},
 volume={27},
 number={11},
 pages={114003},
 year={2018},
 doi = {10.1088/1361-6595/aaec32},
 journal= psst,
 publisher= iop,
 ISSN = {0963-0252} 
}

@article{nava18a,
 title={Experimental Characterization of a 1 k{W} Helicon Plasma Thruster},
 author={Navarro-Cavall{\'e}, J and Wijnen, M and Fajardo, P and Ahedo, E},
 journal={Vacuum},
 volume={149},
 pages={69--73},
 year={2018},
 publisher={Elsevier},
 doi={10.1016/j.vacuum.2017.11.036}
}

@ARTICLE{ramo18a,
  author = jr # and # mm # and # ea,
  title = {Three dimensional fluid-kinetic model of a magnetically guided plasma jet},
  journal = PoP,
  volume = {25},
  number = {6},
  pages = {061206},
  year = {2018},
  ISSN = {1089-7674},
  doi = {10.1063/1.5026972}
}

@Article{sanc18b,
  Title                    = {Kinetic features and non-stationary electron trapping in paraxial magnetic nozzles},
  Author                   = {S{\'a}nchez-Arriaga, Gonzalo and Zhou, Jiewei and Ahedo, E and Mart{\'i}nez-S{\'a}nchez, Manuel and Ramos, Jes{\'u}s Jos{\'e}},
  Journal                  = psst,
  Year                     = {2018},
  Number                   = {3},
  Pages                    = {035002},
  Volume                   = {27},
  Publisher                = iop
}

@article{meri16a,
 author = mm # and # ea,
 title = {Fully magnetized plasma flow in a magnetic nozzle},
 journal = PoP,
 year = {2016},
 volume = {23},
 number = {2},
 pages = {023506},
 numpages = {7},
 publisher = {AIP},
 ISSN = {1089-7674},
 doi = {10.1063/1.4941975}
}

@article{ahed11s,
 author = {Eduardo Ahedo},
 title = {Plasmas for space propulsion},
 journal = {Plasma Physics and Controlled Fusion},
 year = {2011},
 volume = {53},
 pages = {124037},
 number = {12},
 publisher= iop,
 url = {http://stacks.iop.org/0741-3335/53/i=12/a=124037}
}

@article{ahed10f,
 author = ea # and # mm,
 title = {Two-Dimensional Supersonic Plasma Acceleration in a Magnetic Nozzle},
 journal = PoP,
 year = {2010},
 volume = {17},
 number = {7},
 pages = {073501},
 numpages = {15},
 ISSN = {1089-7674},
 publisher = {AIP Publishing, Melville, {NY}},
 doi = {10.1063/1.3442736}
}

@inproceedings{jime24c,
  title={An Implicit Energy- and Charge-conserving Electromagnetic PIC algorithm for Paraxial Magnetic Nozzles},
  author={P. Jim\'enez  and  M. Merino and L. Chac\'on},
  booktitle = iepc38,
  year = {2024},
  number = {IEPC-2024-378},
  address = {Toulouse, France, June 23-28},
  publisher = erps,
}

@InProceedings{inch24b,
  title={Simulation of the discharge and microwave-plasma coupling in a waveguide {ECR} thruster},
  author= mi # and # pj # and # jz # and # mm # and # jn,
  booktitle = iepc38,
  year = {2024},
  number = {IEPC-2024-592},
  address = {Toulouse, France, June 23-28},
  publisher = erps,
}

@InProceedings{mari23a,
  title={Macroscopic response of a {H}all thruster discharge from an axial-radial {PIC} model},
  author= amc # and # ebb # and # adv # and # ea,
  booktitle = gec76,
  year = {2023},
  address = {Ann Arbor, MI, October 9-13}
}

@INPROCEEDINGS{inch21b,
  author = {Marco Inchingolo and Jaume Navarro-Cavall\'e and Mario Merino},
  title = {Design and Plume Characterization of a Low-Power Circular Waveguide Coupled ECR Thruster},
  booktitle = {5$^{th}$ International Workshop on Micropropulsion and CubeSats},
  year = {2021},  
  address = {Toulouse (online)}
}

@incollection{meri16g,
  author = mm # and # ea,
  editor = {J. Leon Shohet},
  title = {Magnetic Nozzles for Space Plasma Thrusters},
  booktitle = {Encyclopedia of Plasma Technology},
  volume = {2},
  publisher = {Taylor and Francis},
  pages = {1329--1351},
  year = {2016},
  ISBN = {9781466500594},
}

@article{chacon2025local,
  title={Local conservation of energy in fully implicit PIC algorithms},
  author={Chac{\'o}n, Luis and Chen, Guangye},
  journal={Journal of Computational Physics},
  volume={529},
  pages={113862},
  year={2025},
  publisher={Elsevier}
}

@article{chacon2013charge,
  title={A charge-and energy-conserving implicit, electrostatic particle-in-cell algorithm on mapped computational meshes},
  author={Chac{\'o}n, Luis and Chen, Guangye and Barnes, Daniel C},
  journal={Journal of Computational Physics},
  volume={233},
  pages={1--9},
  year={2013},
  publisher={Elsevier}
}

@STRING{adv = {Adri\'{a}n Dom\'{i}nguez-V\'{a}zquez}}

@STRING{AIAA = {American Institute of Aeronautics and Astronautics, Reston, {VA}}}

@STRING{erps = {Electric Rocket Propulsion Society, Fairview Park, OH}}

@STRING{sn = {Springer Nature}}

@STRING{iop = {IOP Publishing, Bristol, UK}}

@article{vinci23a,
  title={Enhanced electron heating in the magnetic nozzle of a radio-frequency plasma via electron cyclotron resonance},
  author={Vinci, Alfio E and Mazouffre, St{\'e}phane},
  journal={Europhysics Letters},
  volume={141},
  number={4},
  pages={44002},
  year={2023},
  publisher={IOP Publishing}
}

@article{chen14,
  title={Fluid preconditioning for Newton--Krylov-based, fully implicit, electrostatic particle-in-cell simulations},
  author={Chen, Guangye and Chac{\'o}n, Luis and Leibs, Christopher A and Knoll, Dana A and Taitano, William},
  journal={Journal of computational physics},
  volume={258},
  pages={555--567},
  year={2014},
  publisher={Elsevier}
}

@article{chen14b,
  title={An energy-and charge-conserving, nonlinearly implicit, electromagnetic 1D-3V {V}lasov--{D}arwin particle-in-cell algorithm},
  author={Chen, Guangye and Chac{\'o}n, Luis},
  journal={Computer Physics Communications},
  volume={185},
  number={10},
  pages={2391--2402},
  year={2014},
  publisher={Elsevier}
}

@article{chen15a,
  title={A multi-dimensional, energy-and charge-conserving, nonlinearly implicit, electromagnetic {V}lasov--{D}arwin particle-in-cell algorithm},
  author={Chen, Guangye and Chacon, Luis},
  journal={Computer Physics Communications},
  volume={197},
  pages={73--87},
  year={2015},
  publisher={Elsevier}
}

@article{andr22,
  title={Fully kinetic model of plasma expansion in a magnetic nozzle},
  author={Andrews, Shaun and Di Fede, Simone and Magarotto, Mirko},
  journal={Plasma Sources Science and Technology},
  year={2022},
  publisher={IOP Publishing}
}

@article{barn21,
  title={Finite spatial-grid effects in energy-conserving particle-in-cell algorithms},
  author={Barnes, DC and Chac{\'o}n, Luis},
  journal={Computer Physics Communications},
  volume={258},
  pages={107560},
  year={2021},
  publisher={Elsevier}
}

@article{bath17a,
  title={Electrodeless plasma thrusters for spacecraft: a review},
  author={Bathgate, SN and Bilek, MMM and Mckenzie, DR},
  journal={Plasma Science and Technology},
  volume={19},
  number={8},
  pages={083001},
  year={2017},
  publisher={IOP Publishing}
}

@Book{BIRD91,
  Title                    = {Plasma Physics via Computer Simulation},
  Author                   = {Birdsall, C.K. and Langdon, A.B.},
  Publisher                = {Institute of Physics Publishing},
  Year                     = {1991},

  Address                  = {Bristol}
}

@article{chac16,
  title={A curvilinear, fully implicit, conservative electromagnetic PIC algorithm in multiple dimensions},
  author={Chac{\'o}n, Luis and Chen, Guangye},
  journal={Journal of computational physics},
  volume={316},
  pages={578--597},
  year={2016},
  publisher={Elsevier}
}

@article{carl18,
  doi = {10.1063/1.5017467},
  url = {https://doi.org/10.1063/1.5017467},
  year = {2018},
  publisher = {{AIP} Publishing},
  volume = {25},
  number = {6},
  author = {Johan Carlsson and Igor Kaganovich and Andrew Powis and Yevgeny Raitses and Ivan Romadanov and Andrei Smolyakov},
  title = {Particle-in-cell simulations of anomalous transport in a {P}enning discharge},
  journal = {Physics of Plasmas}
}

@Article{chen11,
  title={An energy-and charge-conserving, implicit, electrostatic particle-in-cell algorithm},
  author={Chen, Guangye and Chac{\'o}n, Luis and Barnes, Daniel C},
  journal={Journal of Computational Physics},
  volume={230},
  number={18},
  pages={7018--7036},
  year={2011},
  publisher={Elsevier}
}

@article{desa23a,
  title={{ECRA} Thruster Advances: 30W and 200W Prototypes Latest Performances},
  author={D{\'e}sangles, Victor and Packan, Denis and Jarrige, Julien and Peterschmitt, Simon and Dietz, Patrick and Scharmann, Steffen and Holste, Kristof and Klar, Peter J},
  journal={Journal of Electric Propulsion},
  volume={2},
  number={1},
  pages={10},
  year={2023},
  publisher={Springer},
  doi ={10.1007/s44205-023-00046-x}
}

@article{erem22,
  title={An energy-and charge-conserving electrostatic implicit particle-in-cell algorithm for simulations of collisional bounded plasmas},
  author={Eremin, Denis},
  journal={Journal of Computational Physics},
  volume={452},
  pages={110934},
  year={2022},
  publisher={Elsevier}
}

@Book{HOCK88,
  Title                    = {Computer simulation using particles},
  Author                   = {Hockney, R.W. and Eastwood, J.W.},
  Publisher                = {CRC Press, Boca Ratón, {FL}},
  Year                     = {1988}
}

@Article{hu17,
  Title                    = {Fully kinetic simulations of collisionless, mesothermal plasma emission: macroscopic plume structure and microscopic electron characteristics},
  Author                   = {Hu, Y. and Wang, J.},
  Journal                  = PoP,
  Year                     = {2017},
  Number                   = {3},
  Pages                    = {033510},
  Volume                   = {24},

  Publisher                = {AIP Publishing, Melville, {NY}}
}

@article{krau07,
	author = {Krause, Todd B. and Apte, A. and Morrison, P.J.},
	title = {A unified approach to the Darwin approximation},
	year = {2007},
	journal = {Physics of Plasmas},
	volume = {14},
	number = {10},
	doi = {10.1063/1.2799346}
}

@Article{kuri70,
  Title                    = {Experimental Study of a Plasma Flow in a Magnetic Nozzle},
  Author                   = {Kuriki, K. and Okada, O.},
  Journal                  = {Physics of Fluids},
  Year                     = {1970},
  Number                   = {9},
  Pages                    = {2262},
  Volume                   = {13}
}

@article{ligh01,
  title={Low frequency electrostatic instability in a helicon plasma},
  author={Light, Max and Chen, Francis F and Colestock, PL},
  journal={Physics of Plasmas},
  volume={8},
  number={10},
  pages={4675--4689},
  year={2001},
  publisher={American Institute of Physics}
}

@article{lape17,
  title={Exactly energy conserving semi-implicit particle in cell formulation},
  author={Lapenta, Giovanni},
  journal={Journal of Computational Physics},
  volume={334},
  pages={349--366},
  year={2017},
  publisher={Elsevier}
}

@article{litt19,
  title={Electron Demagnetization in a Magnetically Expanding Plasma},
  author={Little, Justin M and Choueiri, Edgar Y},
  journal={Physical review letters},
  volume={123},
  number={14},
  pages={145001},
  year={2019},
  publisher={APS}
}

@Article{mart15,
  Title                    = {Electron cooling and finite potential drop in a magnetized plasma expansion},
  Author                   = {Mart{\'i}nez-S{\'a}nchez, M. and Navarro-Cavall{\'e}, J. and Ahedo, E.},
  Journal                  = PoP,
  Year                     = {2015},
  Number                   = {5},
  Pages                    = {053501},
  Volume                   = {22},

  Publisher                = {AIP Publishing, Melville, {NY}}
}

@article{mazo16a,
  title={Electric Propulsion for Satellites and Spacecraft: established Technologies and Novel Approaches},
  author={Mazouffre, St{\'e}phane},
  journal={Plasma Sources Science and Technology},
  volume={25},
  number={3},
  pages={033002},
  year={2016},
  publisher={IOP Publishing},
  DOI={10.1088/0963-0252/25/3/033002}
}

@article{mill66,
  title={Cyclotron resonance thruster design techniques.},
  author={Miller, David B and Bethke, George W},
  journal={AIAA Journal},
  volume={4},
  number={5},
  pages={835--840},
  year={1966}
}

@article{pork72,
  title={Parametric instabilities in a magnetic field and possible applications to heating of plasmas},
  author={Porkolab, Miklos},
  journal={Nuclear fusion},
  volume={12},
  number={3},
  pages={329},
  year={1972},
  publisher={IOP Publishing}
}

@article{port23,
  title={Anisotropic electron heating in an electron cyclotron resonance thruster with magnetic nozzle},
  author={Porto, Jean and Elias, Paul-Quentin and Ciardi, Andrea},
  journal={Physics of Plasmas},
  volume={30},
  number={2},
  year={2023},
  publisher={AIP Publishing}
}

@article{chen23b,
  title={An implicit, conservative and asymptotic-preserving electrostatic particle-in-cell algorithm for arbitrarily magnetized plasmas in uniform magnetic fields},
  author={Chen, Guangye and Chac{\'o}n, Luis},
  journal={Journal of Computational Physics},
  volume={487},
  pages={112160},
  year={2023},
  publisher={Elsevier}
}

@phdthesis{serc93,
  title                     ={An experimental and theoretical study of the {ECR} plasma engine},
  author                    ={Sercel, J.C.},
  year                      ={1993},
  school                    ={California Institute of Technology}
}

@article{shin14a,
  title={Development of electrodeless plasma thrusters with high-density helicon plasma sources},
  author={Shinohara, Shunjiro and Nishida, Hiroyuki and Tanikawa, Takao and Hada, Tohru and Funaki, Ikkoh and Shamrai, Konstantin P},
  journal={IEEE Transactions on Plasma Science},
  volume={42},
  number={5},
  pages={1245--1254},
  year={2014},
  publisher={IEEE}
}

@book{STIX92,
  title={Waves in plasmas},
  author={Stix, Thomas H},
  year={1992},
  publisher={Springer Science \& Business Media}
}

@article{tacc23,
   author = {Taccogna, F. and Cichocki, F. and Eremin, D. and Fubiani, G. and Garrigues, L.},
    title = "{Plasma propulsion modeling with particle-based algorithms}",
    journal = {Journal of Applied Physics},
    volume = {134},
    number = {15},
    pages = {150901},
    year = {2023},
    month = {10},
    issn = {0021-8979},
    doi = {10.1063/5.0153862},
    url = {https://doi.org/10.1063/5.0153862},
}

@Article{taka12,
  Title                    = {Axial force imparted by a current-free magnetically expanding plasma},
  Author                   = {Takahashi, Kazunori and Lafleur, Trevor and Charles, Christine and Alexander, Peter and Boswell, Rod W},
  Journal                  = PoP,
  Year                     = {2012},
  Number                   = {8},
  Pages                    = {083509},
  Volume                   = {19}
}

@article{taka13b,
  title={Performance improvement of a permanent magnet helicon plasma thruster},
  author={Takahashi, Kazunori and Charles, Christine and Boswell, Rod and Ando, Akira},
  journal={Journal of Physics D: Applied Physics},
  volume={46},
  number={35},
  pages={352001},
  year={2013},
  publisher={IOP Publishing}
}

@article{taka19,
  title={Helicon--Type Radiofrequency Plasma Thrusters and Magnetic Plasma Nozzles},
  author={K. Takahashi},
  journal={Reviews of Modern Plasma Physics},
  volume={3},
  number={},
  pages={3},
  year={2019},
  publisher={Springer},
  doi={10.1007/s41614-019-0024-2}
}

@inproceedings{hewe85,
  title={Elimination of electromagnetic radiation in plasma simulation: The {D}arwin or magneto inductive approximation},
  author={Hewett, Dennis W},
  booktitle={Space Plasma Simulations: Proceedings of the Second International School for Space Simulations, Kapaa, Hawaii, February 4--15, 1985},
  pages={29--40},
  year={1985},
  organization={Springer}
}

@article{huangFiniteGridInstability2016,
	title = {Finite grid instability and spectral fidelity of the electrostatic {Particle}-{In}-{Cell} algorithm},
	volume = {207},
	issn = {0010-4655},
	url = {https://www.sciencedirect.com/science/article/pii/S001046551630145X},
	doi = {10.1016/j.cpc.2016.05.021},
	urldate = {2025-01-07},
	journal = {Computer Physics Communications},
	author = {Huang, C. -K. and Zeng, Y. and Wang, Y. and Meyers, M. D. and Yi, S. and Albright, B. J.},
	month = oct,
	year = {2016},
	pages = {123--135},
}

\end{document}